\documentclass[journal]{IEEEtran}
\usepackage{amsmath,amsfonts}
\usepackage{algorithmic}
\usepackage{algorithm}
\usepackage{multirow}
\usepackage{array}
\usepackage[caption=false,font=normalsize,labelfont=sf,textfont=sf]{subfig}
\usepackage{textcomp}
\usepackage{stfloats}
\usepackage{url}
\usepackage{svg}
\usepackage{hyperref}
\usepackage{verbatim}
\usepackage{graphicx}
\usepackage{cite}
\usepackage{xcolor}
\usepackage{svg}
\usepackage{microtype}
\IEEEoverridecommandlockouts

\usepackage{booktabs}
\begin{document}

\title{LiDAR-Guided Cross-Attention Fusion for Hyperspectral  Band Selection and Image Classification}

\author{{Judy X~Yang,~\IEEEmembership{Student Member,~IEEE},
        Jun~Zhou,~\IEEEmembership{Senior Member,~IEEE},
        Jing~Wang,~\IEEEmembership{Senior Member,~IEEE},
        Hui~Tian,~\IEEEmembership{Senior Member,~IEEE}, and~Alan Wee-Chung~Liew,~\IEEEmembership{Senior Member,~IEEE}}  
        
\thanks{Judy X Yang, Jun Zhou, Hui Tian and Alan Wee-Chung Liew are with the School of Information and Communication Technology, Griffith University, Australia (corresponding author: Jun Zhou, jun.zhou@griffith.edu.au).}
\thanks{Jing Wang is with the Queensland Department of Agriculture and Fisheries, Australia.}}



\maketitle

\begin{abstract}
The fusion of hyperspectral and LiDAR data has been an active research topic. Existing fusion methods have ignored the high-dimensionality and redundancy challenges in hyperspectral images, despite that band selection methods have been intensively studied for hyperspectral image (HSI) processing. This paper addresses this significant gap by introducing a cross-attention mechanism from the transformer architecture for the selection of HSI bands guided by LiDAR data. LiDAR provides high-resolution vertical structural information, which can be useful in distinguishing different types of land cover that may have similar spectral signatures but different structural profiles. In our approach, the LiDAR data are used as the ``query" to search and identify the ``key" from the HSI to choose the most pertinent bands for LiDAR. This method ensures that the selected HSI bands drastically reduce redundancy and computational requirements while working optimally with the LiDAR data. Extensive experiments have been undertaken on three paired HSI and LiDAR data sets: Houston 2013, Trento and MUUFL. The results highlight the superiority of the cross-attention mechanism, underlining the enhanced classification accuracy of the identified HSI bands when fused with the LiDAR features. The results also show that the use of fewer bands combined with LiDAR surpasses the performance of state-of-the-art fusion models.

\end{abstract}

\begin{IEEEkeywords}
Remote sensing, multimodal data fusion, hyperspectral images, LiDAR data, band selection, cross-attention.
\end{IEEEkeywords}

\section{Introduction} 
Remote sensing has witnessed remarkable advances in recent decades and generated significant impacts in agriculture, environmental monitoring, and urban planning~\cite{vadrevu2022remote}. An active research topic in remote sensing is multimodal data fusion, which aims to explore heterogeneous features extracted from various sources and achieves results that single source data cannot achieve on its own~\cite{lai2022review}.

Among data captured by various sensors, hyperspectral image (HSI) and light detection and range (LiDAR) data have unique advantages. HSI sensors capture the Earth's surface in a wide range of electromagnetic spectrum, offering detailed spectral information for material identification. LiDAR data, on the other hand, are capable of providing 3D profiles to measure the distance between the sensor and the surface of the Earth, offering precise topographic and structural information~\cite{gao2020survey}. Fusion of hyperspectral and LiDAR data can leverage the complementary advantages of both data modalities and lead to improved image classification accuracy~\cite{10229170}.

The rich spectral detail provided by the HSI results in high dimensionality and redundancy in HSI data, which leads to challenges such as the Hughes phenomenon, a large computational load, and inefficiency in storage and transmission. Hyperspectral band selection is one way to mitigate these challenges, which attempts to identify and preserve a subset of spectral bands from the original HSI that contribute the most pertinent information while discarding redundant or irrelevant bands. Hyperspectral band selection has been widely studied, whose methods fall mainly into five categories: ranking-based, search-based,
clustering-based, deep learning-based, and their hybrid ~\cite{sun2019hyperspectral}. 

In parallel, significant research efforts have been dedicated to the fusion of hyperspectral and LiDAR data to improve the classification of land cover. Methods on this topic can be broadly categorised into traditional~\cite{jahan2018fusion} and deep learning-based approaches ~\cite{khan2021deep,falahatnejad2022deep}. Traditional fusion methods typically involve pixel-based or feature-based approaches, using statistical and signal processing techniques such as PCA or wavelet transform to combine and extract valuable features from data sets \cite{1433035,kaur2021image,liu2017feature}. On the other hand, deep learning-based approaches have introduced deep neural networks \cite{li2019deep,hong2020more}  and vision transformers for joint hyperspectral and LiDAR classification, which address limitations in heterogeneous feature representation and information fusion~\cite{xue2022deep,zhang2022convolution}.

\begin{figure*}[t]
\centering
\subfloat[]{\includegraphics[width=6cm, height=3cm]{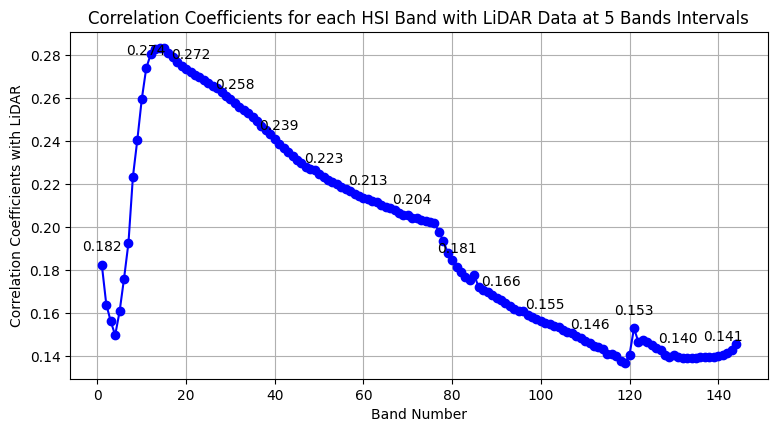}%
\label{fig_uh_cr}}
\hfil
\subfloat[]{\includegraphics[width=6cm, height=3cm]{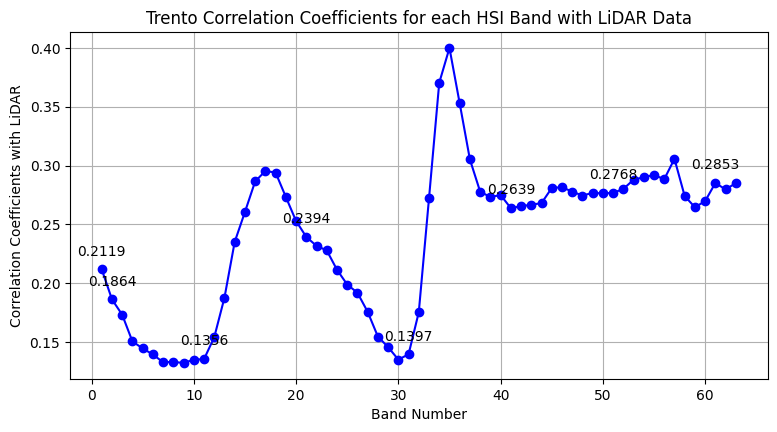}%
\label{fig_tr_cr}}
\hfil
\subfloat[]{\includegraphics[width=6cm, height=3cm]{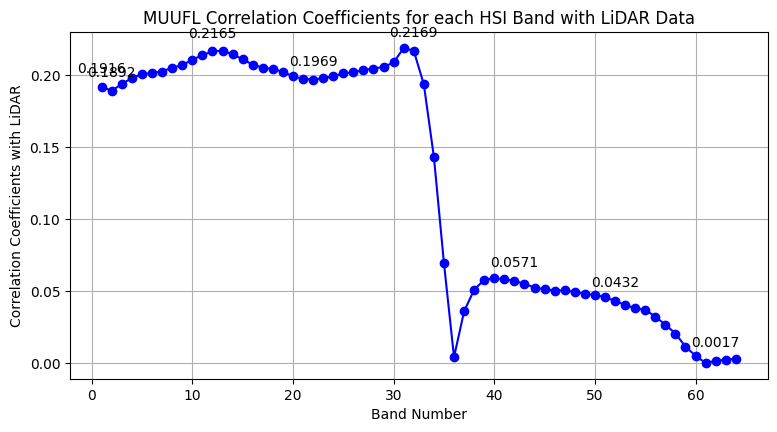}%
\label{fig_mf_cr}}
\caption{{Correlation coefficients between LiDAR and each hyperspectral band of Three Datasets. (a)  Houston2013 HSI and LiDAR. (b) Trento HSI and LiDAR. (c) MUUFLE HSI and LIDAR.  The horizontal axis shows the band index.}}
\label{figcc}
\end{figure*}

Fig.~\ref{figcc}  {shows the correlation coefficients between LiDAR and each hyperspectral band of
three datasets: Houston 2013, Trento, and MUUFL.~\cite{6776408, university_of_trento_theses_2022,du_zare_2017}. We can see that a certain level of correlation between LiDAR and hyperspectral bands exists. This correlation fluctuates among different spectral bands. It is generally believed that LiDAR data contain distance information, and HSI contain chemical components information, which seems to be independent and complementary. However, two factors could contribute to such correlation phenomenon.
First, the characteristics of the ground cover structures may cause the correlation between LiDAR and HSI bands. For example, tree canopy and ground soil in a same area have difference heights, which can be captured by LiDAR. Meanwhile, these two classed have different chemical components, and thus have different spectral responses. This leads to a certain
level of correlation between LiDAR and HSI. Since different spectral bands are sensitive to different materials, we can see that the correlation varies across different bands and different datasets. Second, LiDAR also relies on the specific wavelength of the laser to sense the objects. The relationship of wavelength in LiDAR and HSI bands could have contribution to the correlation phenomenon}

{Additonally}, HSI sensors produce 2D raster-based images that have a low spatial resolution compared to LiDAR data. LiDAR sensors produce 3D. Although there is no direct correspondence one-to-one between the two datasets~\cite{kim2016integrated,hong2020more}, it can be seen that some hyperspectral bands have correlations with the LiDAR data more than others. In addition to containing complementary information, some hyperspectral bands, after adding the LiDAR information into the raster-based HSI process, helped to distinguish spectrally similar materials, such as trees and grass~\cite{kim2016integrated}. When selecting important hyperspectral bands with LiDAR, we also need to consider the overlap of this information. The importance level of a hyperspectral band decreases when it is highly correlated with the already available LiDAR. In statistics, Pearson's correlation coefficient measures the linear relationship between two sets of data. A higher coefficient indicates a stronger linear correlation. Therefore, when the coefficient is high, it suggests that the data points of the two data sets (in this case, LiDAR and certain hyperspectral bands) have a strong linear relationship, implying that they contain similar information~\cite{poulton2019long, qamar2023atmospheric}

The integration of LiDAR data into raster-based HSI analysis increases the distinction of spectrally similar materials, exemplified by the differentiation of trees and grass~\cite{kim2016integrated}. Therefore, it is imperative to consider the informational overlap when selecting pertinent HSI bands to pair with LiDAR data. The utility of an HSI band may diminish if its correlation with existing LiDAR data is substantial. Pearson's correlation coefficient, as a statistical measure, gauges the linearity of the relationship between two datasets. A pronounced coefficient suggests a robust linear association, signifying considerable overlap in the information conveyed by the respective datasets. Thus, the high correlation coefficients between LiDAR and certain HSI bands underscore the presence of analogous information within these paired datasets.


Building on the hyperspectral band selection and HSI and LiDAR fusion techniques, we identify the following two research gaps in the existing literature. First, there is a lack of well-defined methods for selecting the most pertinent bands in HSI when LiDAR data are available, limiting the potential synergies that can be exploited by integrating LiDAR and HSI data for enhanced data classification and analysis. Although traditional band selection methods can select important bands from hyperspectral data, they are designed for hyperspectral data only. When paired with LiDAR data, the selected band of these methods is not guaranteed to be optimal.

Second, although it is widely accepted that the Hughes phenomenon hampers the performance of hyperspectral image classification, to the best of our knowledge, almost all hyperspectral and LiDAR fusion methods still use all hyperspectral bands. It is important to investigate whether the fusion of fewer bands with LiDAR would lead to better classification performance. 

We strive to bridge these gaps in current research on band selection and data fusion. Inspired by the attention mechanism of transformer architectures in domains such as natural language processing and image processing, we propose a novel paradigm for HSI band selection, guided by the attributes of LiDAR data. We aim to show that leveraging LiDAR's spatial properties with HSI's spectral properties can lead to a new solution in HSI and LiDAR data fusion.
 
The contributions of this paper are summarised as follows:
\begin{enumerate}
\item We identify the existing gaps in HSI and LiDAR data fusion and band selection research. Then we leverage the latest success of transformer architecture and propose a novel cross-attention mechanism that uses LiDAR data to guide HSI band selection.
\item Our model is tested on three HSI and LiDAR data fusion datasets and consistently outperforms existing band selection techniques in HSI classification. The experiments demonstrate the effectiveness of our model.
\item Our approach demonstrates that selecting significant bands and combining them with LiDAR data can achieve better performance than current fusion models, suggesting new prospects for remote sensing data fusion.
\end{enumerate}

The paper is structured as follows. Section~\ref{sec:relatedwork} reviews related works in the literature. Our method is described in Section~\ref{sec:methodology}, followed by the experiments and results presented in Section~\ref{sec:experiments}. Finally, Section~\ref{sec:conclusions} concludes this article and highlights directions for future research.

\section{Related Work}~\label{sec:relatedwork}
HSI and LiDAR data fusion has shown its effectiveness in improving the performance of various remote sensing applications, such as land cover classification, target detection, and environmental monitoring. In this section, we review the literature with a focus on band selection strategies and data fusion methodologies relevant to our study.

\subsection{Hyperspectral Band Selection}
The core objective of hyperspectral band selection is to identify crucial and distinguishing bands in HSI, which aims to reduce data dimensionality and computational cost while preserving vital information in the images. We briefly review five predominant categories of hyperspectral band selection methods: ranking-based, searching-based, clustering-based, deep learning-based, and hybrid methods~\cite{sun2019hyperspectral,sawant2021band}.

Ranking-based methods prioritise band indexes using different statistical data analyses, for example, covariance~\cite{kim2017covariance}, correlation~\cite{9528953}, and structural similarity~\cite{xu2021similarity}. These approaches rank the bands based on the uniqueness or the representative of the bands on a given dataset. Searching-based methods perform band selection using a distance measurement function with a searching strategy.    Distance measurements used in hyperspectral imaging are crucial for various tasks. Spectral Angle Mapping (SAM), as mentioned in Liu et al. (2013)\cite{zhang2022spatial}, quantifies uncertainty or randomness in spectral information, useful for assessing data complexity. The Maximum Ellipsoid Volume, as discussed in~\cite{8320544}, is used to maximise the data scatter in a reduced-dimensional space. The Structural Similarity Index (SSI)~\cite{xu2021similarity} is another important measure for assessing the similarity between images or spectra.
The categorisation of search strategies in hyperspectral band selection, as described by Sun et al. (2019)~\cite{sun2019hyperspectral}, includes incremental, updated and eliminating strategies, each with a specific methodology for selecting the optimal bands. Incremental strategy iteratively adds bands, updated strategy refines choices as new data become available, while eliminating strategy removes less informative bands systematically.

Clustering-based methods group the original bands into clusters, selecting representative bands from each for the final subset. Initially using hierarchical clustering with Ward linkage~\cite{zhang2022spatial,wang2022region} , subsequent methods used information measurements such as MI\cite{cao2016automatic} or Kullback-Leibler divergence\cite{zhang2022spatial}. Commonly used clustering approaches consist of k-means~\cite{wang2022hyperspectral}, affinity propagation~\cite{tahraoui2017affinity} , graph clustering~\cite{cai2020graph} , optimal band clustering~\cite{wang2018optimal}, fuzzy clustering~\cite{zhang2017unsupervised}, and spectral-spatial subspace clustering~\cite{wang2022graph}. 

Deep learning-based methods adopt deep neural networks for band selection. Among them, BSNets~\cite{cai2019bs} adopt the Band Attention Module (BAM) to model complex relationships between spectral bands, then use a Reconstruction Network (RecNet) to validate the chosen bands by reconstructing the original HSI from the BAM-selected bands. Deep reinforcement learning~\cite{9387453}  frames unsupervised band selection as a Markov decision process and automatically learns a policy to select an optimal subset of bands. Constrained measurement learning network~\cite{ayna2023learning} is composed of a constrained measurement learning network for band selection, followed by a classification network. These two networks are jointly trained to minimise the classification loss along band selection. Deep learning with spatial features~\cite{sawant2021band} integrates spectral-spatial features into a deep learning framework for the selection of hyperspectral bands. These methods provide a variety of ways to select informative and non-redundant bands from hyperspectral data using deep learning.

Hybrid band-selection methods employ multiple strategies for optimal band selection. Commonly, they combine clustering and ranking approaches~\cite{yu2021semisupervised}. Zhao et al. presented an unsupervised Spectral-Spatial genetic algorithm-based method~\cite{9321482}, which integrated search-based, deep learning-based, rank-based, and clustering-based approaches. The spectral separability index algorithm integrates clustering, classification and searching for the best band combination~\cite{9321482,chang2014hyperspectral,wang2016salient}. These hybrids aim to achieve higher classification accuracy and band significance by leveraging the strengths of multiple selection techniques.

Although current band-selection studies predominantly focus on HSI data, our objective is to identify the most informative hyperspectral bands in the presence of LiDAR data, and we aim to compare our method with existing band-selection approaches using the same paired HSI and LiDAR data sets. 

\subsection{HSI and LiDAR Fusion Models}
HSI and LiDAR data fusion has witnessed substantial progress. Although traditional methods have their merits~\cite{jahan2020inverse}, deep learning techniques have proven superior in handling the complexity of these data sets~\cite{hong2020more}. In particular, convolutional neural network (CNN)-based architectures have been instrumental in HSI and LiDAR data fusion, offering robust spatial feature extraction capabilities~\cite{8985546,falahatnejad2022deep}. Huang et al.~\cite{8985546} introduced a framework that uses two coupled convolutional neural networks (CNN) to fuse hyperspectral and LiDAR data, extracting spectral-spatial features and elevation information, respectively. EndNet~\cite{hong2020deep}, an Encoder-Decoder fusion network, addresses the limitations of single-modal remote sensing data and uses a reconstruction strategy to activate neurons in both modalities. The collaborative contrastive learning (CCL) method~\cite{jia2023collaborative} uses a pretraining and fine-tuning strategy to extract features from HSI and LiDAR data separately and produce coordinated representation and matching of features between the two-modal data without label samples. Integrating single-band and multi-band LiDAR fusion with hyperspectral data, a deep learning fusion framework~\cite{8529194} was introduced to merge HSI and LiDAR for tree species mapping. 

With the integration of deep learning and attention-based mechanisms~\cite{zhang2016deep}, the effectiveness of data fusion becomes more effective. The FusAtNet model~\cite{mohla2020fusatnet} fuses self-attention and cross-attention {module}, highlighting the harmonious integration of the HSI and LiDAR data. Falahatnejad et al.~\cite{falahatnejad2022deep} used hybrid 3D/2D CNNs and attention {module} to extract spatial-spectral characteristics of HSI and representative elevation characteristics of LiDAR. The extracted features were then integrated and classified using a softmax classifier. The collaborative attention network~\cite{zhou2022joint} addresses limitations in the representation of heterogeneous features. Both multiscale and single-branch networks are used to extract land cover features enhanced by self-attention and cross-attention {module}. 

Deep learning and attention-based fusion techniques take advantage of the strengths of both types of data, providing comprehensive insight that is crucial for various applications. However, there is a noticeable gap as current approaches tend to use full HSI bands and LiDAR data without performing any band selection. This lack of band selection can lead to unnecessary computational complexity and potential information redundancy, highlighting the need for further research and development in this area.

\section{Methodology}\label{sec:methodology}
This section introduces the proposed integrated LiDAR-guided band selection framework. This framework consists mainly of two modules: the self-attention module for feature extraction from hyperspectral and LiDAR data and the LiDAR-guided cross-attention module. We first introduce the overview framework, followed by the details of the attention {module}.

\subsection{Overview of the Proposed Fusion Framework}

Our method leverages the transformer, originally prominent in natural language processing (NLP) for sequence-to-sequence tasks such as machine translation~\cite{vaswani2017attention}, for HSI and LiDAR data fusion. Transformers have revolutionised data processing by using self-attention mechanisms to model long-range dependency in the data. Beyond NLP, the potential of transformers has been validated in computer vision and hyperspectral image classification, with innovations such as the Vision Transformer (ViT)~\cite{dosovitskiy2020image} achieving or even surpassing traditional CNN benchmarks in various visual tasks. 

\begin{figure*}[t]
\centerline{\includegraphics[width=19cm, height=6cm]{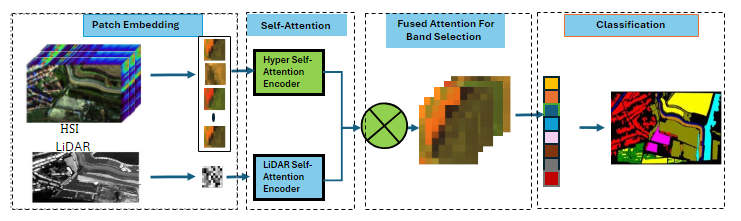}}
\caption{The overall framework of the proposed LiDAR-guided cross-attention fusion for band selection.}
\label{fig1_m_v2}.
\end{figure*}

\begin{figure}[t]
\centerline{\includegraphics[width=8cm, height=12cm]{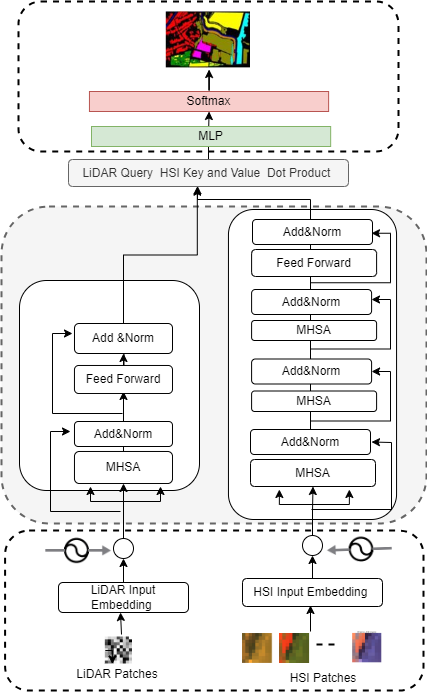}}
\caption{The detailed illustration of the  the Proposed LiDAR-guided cross-attention fusion for band selection method. The network consists of four components: A) input and patch embedding module; B) self-attention module; C)cross-attention module; and D) MLP.}
\label{ca}.
\end{figure}

Drawing from these advances, we propose an attention-based mechanism adapted from transformer architecture to model the relationship between HSI and LiDAR and obtain the significance of each HSI band in the context of LiDAR fusion. Fig.~\ref{fig1_m_v2} shows the model work flow.  The model consists mainly of four modules,in Figure \ref{ca}, i.e. the input and patch embedding module, the self-attention module, the cross-attention module, and the Multi-Layer Perceptron (MLP) and classification module.  

The input to our model is a hyperspectral image $\mathbf{X_{H}} \in 
\mathbb{R}^{W \times H \times B}$ and the corresponding LiDAR data 
$\mathbf{X_{L}} \in \mathbb{R}^{W \times H \times C}$, where $W$ and $H$ 
represent the width and height of the dataset, and $B$ and $C$ are the 
number of HSI bands and the number of LiDAR channels, respectively. A 
pair of input sample patches extracted from hyperspectral and LiDAR data 
is denoted as $\mathbf{X_{h}} \in \mathbb{R}^{P \times P \times B}$ and 
$\mathbf{X_{l}} \in \mathbb{R}^{P \times P \times C}$, where $P$ 
represents the size of the sample patch. To formulate the raw input data 
into sequences for the transformer-like architecture, we flatten the 
spatial dimensions of the HSI and LiDAR data so that they are represented 
as vectors, with each flattened band representing a token. Thus, the
results are $\mathbf{\hat{X}}_{h0} \in \mathbb{R}^{M \times B}$ and 
$\mathbf{\hat{X}}_{l0} \in \mathbb{R}^{M \times C}$, respectively, where
$M = P^2$.

Although grouping neighbouring bands into one token is often adopted in the hyperspectral image classification~\cite{hong2021spectralformer,hong2023spectralgpt},  we use a band-wise token. The reason is that we can directly assess the importance and relationships of individual bands from hyperspectral and LiDAR data, aligning with our goal of selecting LiDAR-guided hyperspectral bands. Like classical transformers, the inputs of both hyperspectral and LiDAR data go through an embedding layer and are added with positional encoding to keep the band-order information. After embedding, the characteristic of the hyperspectral data becomes $(d,B)$, where $d$ denotes the embedding size. 

The main part of the proposed method comprises two modules: the self-attention module and the cross-attention module. The hyperspectral and LiDAR features go through two separate branches of stacked self-attention \textcolor{blue}{module}. These stacked self-attention modules extract feature representations of each hyperspectral band and LiDAR channel individually. Finally, the two streams interact with each other at the end of the network with a cross-attention module to fuse and weigh the importance of each hyperspectral band with respect to the LiDAR data. The motivation of this design is aligned with our goal of selecting hyperspectral bands using LiDAR as a query. Finally, at the end of the network, the fused and weighted features are sent through the fully connected layer and the softmax function to output the classification results.

\subsection{Self attention and cross attention modules}

Multihead attention was first proposed in language models to process long sequences~\cite{vaswani2017attention}. We adopted it here to learn the representation of each band while considering the context of information within each data modality. Specifically, each hyperspectral band representation $r_{i}$, ${i=1,2,...B}$ is mapped to three spaces, namely ${q_{i}}$, ${k_{i}}$, and ${v_{i}}$. The three spaces correspond to the query, key, and value space. Then the attention can be achieved by:

\begin{equation}
    Attention(\mathbf{Q},\mathbf{K},\mathbf{V}) = softmax(\frac{\mathbf{Q} \cdot \mathbf{K}^T} {\sqrt{d}}) \mathbf V
    \label{selfatt}
\end{equation}

Equation~(\ref{selfatt}) denotes a process of self-attention for a single head. In practical applications, it is beneficial to map the representations to different Query, Key, and Values spaces and perform attention in multiple different spaces. Finally, the results of multiple attentions are concatenated together. A typical transformer encoder layer often has a feed-forward network attached after multi-head attention to learn better features.

\begin{figure}[htbp]
\centering
\includegraphics[width=8cm, height=9cm]{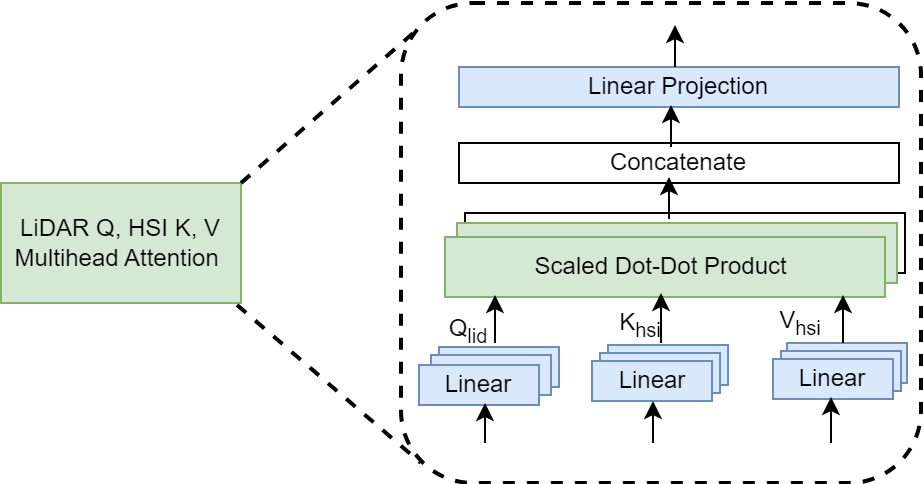}
\caption{Cross attention from LiDAR and HSI: LiDAR queries undergo linear transformation, interact with hyperspectral keys/values through scaled dot-product attention, concatenate across multiple heads, followed by a final linear projection for enriched data fusion.}
\label{fig2_m}
\end{figure}

The stacked self-attention layers can learn highly distinctive features for hyperspectral and LiDAR, respectively, but cannot model the relationship and significance of each hyperspectral band with respect to LiDAR. To achieve this goal, we further design a cross-attention module that facilitates the LiDAR data to focus on specific hyperspectral bands, as shown in Fig.~\ref{fig2_m}.


After going through the stacked self-attention module, the hyperspectral and LiDAR feature representation can be denoted by $\hat{X_{h}}$ and $\hat{X_{l}}$ with sizes $(d,B)$ and $(d,C)$. To model the relationship of LiDAR to each hyperspectral band, we choose a different querying strategy from the self-attention module. Specifically, we only map LiDAR data to the query space and hyperspectral bands to the key space. Then $Q_{L}$ is used to query the hyperspectral band's $K_{H}$. In this manner, the LiDAR and each hyperspectral band can interact directly with each other to model the importance of each hyperspectral band to the LiDAR, such that:
\begin{equation}
   \mathbf{Q}_{L} = \mathbf{W}_q \times \mathbf{\hat{X}}_l, 
\end{equation}
\begin{equation}
\mathbf{K}_{H} = \mathbf{W}_k \times \mathbf{\hat{X}}_h, 
\end{equation}
\begin{equation}
 \mathbf{V}_{H} =\mathbf{W}_v \times \mathbf{\hat{X}}_h.
\end{equation}
where $\mathbf{W}_q, \mathbf{W}_k$ and $\mathbf{W}_v$ are the projection matrices for the three spaces, respectively. Cross-attention weights can be obtained by:
\begin{equation}
    AttWeights_{singlehead} = softmax(\frac{\mathbf{Q_{L}} \cdot \mathbf{K_{H}}^T} {\sqrt{d}})
\end{equation}

For example,  when LiDAR has one channel, hyperspectral has 144 bands, and the embedding dimension is 256, $Q_{L}$ and $K_{H}^{T}$ are in the size of $1 \times 256$ and $256 \times 144$, respectively. Thus, the final result has the size of $1 \times 144$. After applying the softmax function, we can obtain a normalised 144-dimensional weight vector, each element of which represents the relevance of the individual bands of hyperspectral in regards to LiDAR.  The final attention weights are averaged over all heads when adopting multi-head in the cross-attention module:


\begin{equation}
AttWeights_{\text{multihead}} = \sum_{i=1}^{N} \frac{1}{N}\text{softmax}\left(\frac{\mathbf{Q}^{i}_{L} \cdot (\mathbf{K}^{i}_{H})^T}{\sqrt{d_k}}\right)
\end{equation}
where $N$ is the number of heads. 

By applying the attention weights on the hyperspectral band, we can obtain the output of the cross-attention module.
The cross-attention module performs band weighting and data fusion at the same time. Training this whole network leads to the generation of the optimal weights of hyperspectral bands for LiDAR data on the fusion task. Thus, attention weights can be used to select the most useful hyperspectral bands for fusion with LiDAR, i.e., the bands with the highest weights.

Note that the learned attention weights depend on each input pair of hyperspectral and LiDAR samples. This means that the best hyperspectral band subsets for LiDAR fusion might be different for different ground objects. This also makes sense intuitively. On the contrary, we can also obtain a consistent set of selected bands for all classes. A simple solution is to input all training samples into the network, obtain attention weights for each sample, and then average over them. 

\section{Experiments}\label{sec:experiments}
We have undertaken extensive experiments to assess the performance of the proposed cross-attention band selection method. The experiments were run on three paired HSI and LiDAR datasets, Houston 2013, Trento, and MUUFL.

\begin{figure}[t]
\centerline{\includegraphics[width=9cm, height=6cm]{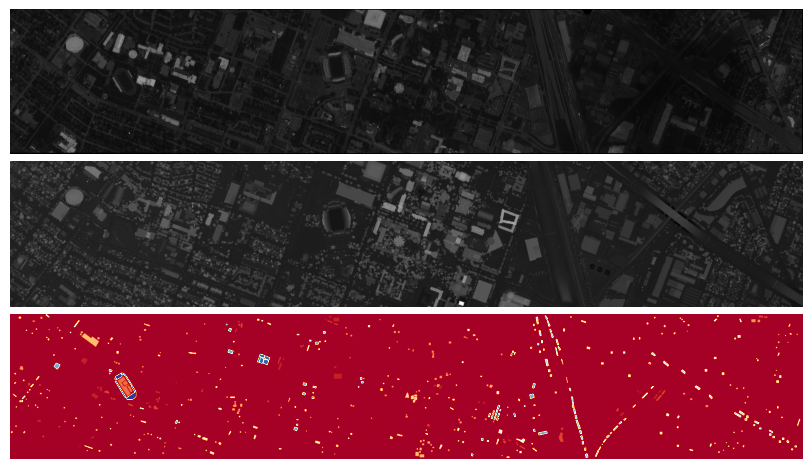}}
\caption{Visualisation of the Houston 2013 data set. Top: HSI; Middle: LiDAR data; Bottom: ground truth.}
\label{fig3}
\end{figure}

{\bf Houston 2013 Dataset:} This dataset contains hyperspectral and LiDAR data for the 2013 IEEE GRSS Data Fusion~\cite{houston_dataset_2013}. It includes a hyperspectral image (HSI) with 144 spectral bands (380-1050nm) and a LiDAR-derived digital surface model (DSM), both at 2.5 m resolution. The HSI is calibrated to the radiance of the sensor, while the DSM indicates elevation in meters above sea level. The data set contains 15 types of land cover, making it ideal for evaluating band selection and classification methods, despite urban complexity and challenges related to noise in the HSI. Details on the training and test samples from the experiment are given in TABLE~\ref{tab:hs2013samples}. Fig.~\ref{fig3} shows the false colour images of the Houston 2013 data set.

\begin{figure}[t]
\centerline{\includegraphics[width=9cm, height=7cm]{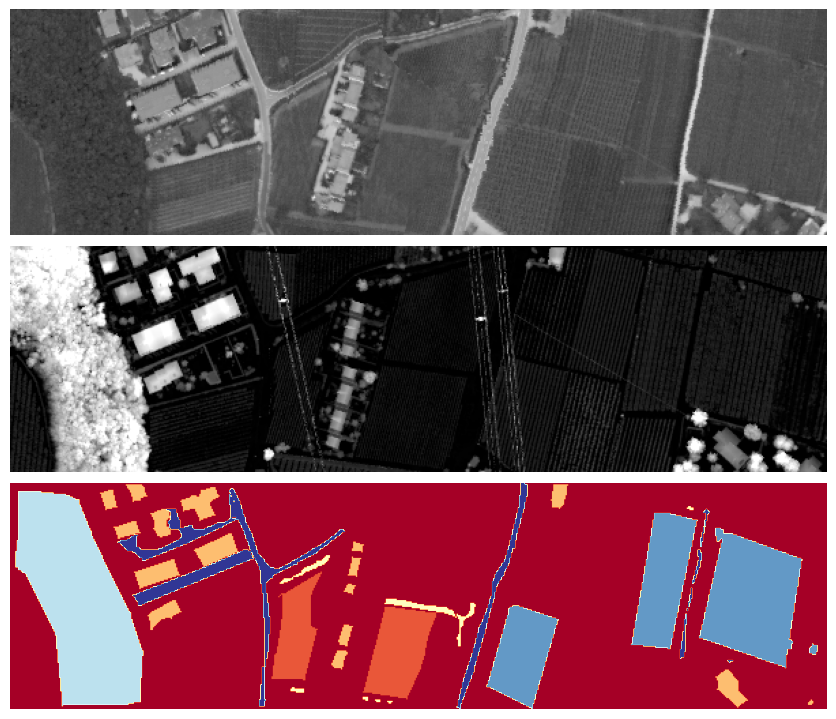}}
\caption{Visualisation of the Trento data set. Top: HSI; Middle: LiDAR data; Bottom: ground truth.}
\label{fig4}
\end{figure}

\begin{figure}[t]
\centerline{\includegraphics[width=9cm, height=4cm]{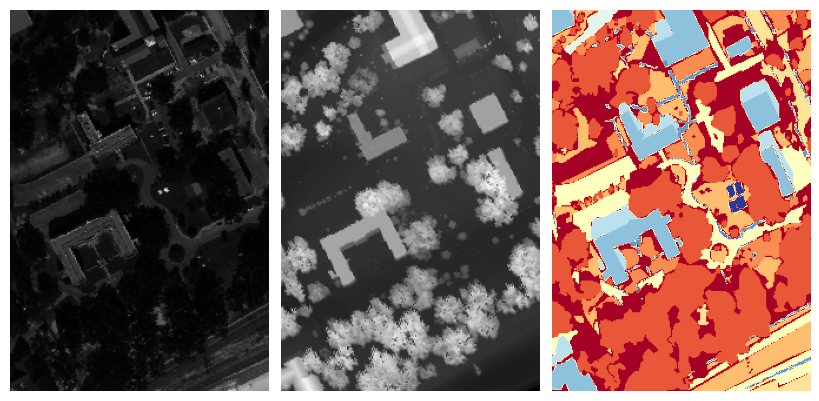}}
\caption{Visualisation of the MUUFL data set. Left: HSI; centre: LiDAR data; right: ground truth.}
\label{fig5}
\end{figure}

\begin{table}[t]
\centering
\caption{Houston 2013 dataset}
\label{tab:hs2013samples}
\fontsize{8}{10}\selectfont
\begin{tabular}{|c||l||c||c||c|}
\hline
\textbf{No.} & \textbf{Class Name} & \textbf{Training} & \textbf{Test} & \textbf{Samples} \\ \hline
1 & Healthy grass & 198 & 1053 & 1251 \\ 
2 & Stressed grass & 190 & 1064 & 1254 \\ 
3 & Synthetic grass & 192 & 505 & 697 \\ 
4 & Tree & 188 & 1056 & 1244 \\ 
5 & Soil & 186 & 1056 & 1242 \\ 
6 & Water & 182 & 143 & 325 \\ 
7 & Residential & 196 & 1072 & 1268 \\ 
8 & Commercial & 191 & 1053 & 1244 \\ 
9 & Road & 193 & 1059 & 1252 \\ 
10 & Highway & 191 & 1036 & 1227 \\ 
11 & Railway & 181 & 1054 & 1235 \\ 
12 & Parking lot 1 & 192 & 1041 & 1233 \\ 
13 & Parking lot 2 & 184 & 285 & 469 \\ 
14 & Tennis court & 181 & 247 & 428 \\ 
15 & Running track & 187 & 473 & 660 \\ 
\hline 
 &\textbf Total  &2832 & 12197 & 15029 \\ 
 \hline 
\end{tabular}
\end{table}

{\bf Trento Data Set:} The Trento data set~\cite{university_of_trento_theses_2022}  was acquired in Trento, Italy, in 2015. The hyperspectral image comprises 48 spectral bands (400-950nm) with one-metre spatial resolution, while the LiDAR data presents the DSM. The data set covers approximately 100 hectares and includes six land cover classes. Table~\ref{tab:Trentosamples} outlines the number of training and test samples used in the current experiment for each class of interest. Fig.~\ref{fig4} shows the false colour images of the Trento data set.

\begin{table}[t]
\centering
\caption{Trento data set}
\label{tab:Trentosamples}
\fontsize{8}{10}\selectfont
\begin{tabular}{|c||l||c||c||c|}
\hline
\textbf{No.} & \textbf{Class Name} & \textbf{Training} & \textbf{Test} & \textbf{Samples} \\ \hline 
1 & Apple trees & 129 & 3905 & 4034 \\
2 & Buildings & 125 & 2778 & 2903 \\
3 & Ground & 105 & 374 & 479 \\
4 & Wood & 154 & 8969 & 9123 \\
5 & Vineyard & 184 & 10,317 & 10501 \\
6 & Roads & 122 & 3052 & 3174 \\
\hline & Total & 819 & 29,395 & 30,214 \\
\hline 
\end{tabular}
\end{table}

{\bf MUUFL Gulfport scene:} The MUUFL Gulfport scene\cite{du_zare_2017}contains hyperspectral and LiDAR data collected over Gulfport, Mississippi, in 2010 and 2011. The data set includes four sub-images with different spatial resolutions and elevations and ground-truth information. TABLE~\ref{tab:MUUFLsamples} gives the distribution of the sample in this experiment. Fig.~\ref{fig5} shows the false-colour images of the MUUFL data set.

\begin{table}[t]
\centering
\caption{MUUFL Data Set}
\label{tab:MUUFLsamples}
\fontsize{8}{10}\selectfont
\begin{tabular}{|c||l||c||c||c|}
\hline
\textbf{No.} & \textbf{Class Name} & \textbf{Training} & \textbf{Test} & 
\textbf{Samples} \\ \hline 
1 & Trees & 150 & 23,246 & 23,396 \\
2 & Mostly grass & 150 & 4,270 & 4,420 \\
3 & Mixed ground surface & 150 & 6,882 & 7,032 \\
4 & Dirt and sand & 150 & 1,826 & 1,976 \\
5 & Road & 150 & 6,687 & 6,837 \\
6 & Water & 150 & 466 & 616 \\
7 & Building shadow & 150 & 2,233 & 2,383 \\
8 & Building & 150 & 6,240 & 6,396 \\
9 & Sidewalk & 150 & 1,385 & 1,535 \\
10 & Yellow curb & 150 & 183 & 333 \\
11 & Cloth panels & 150 & 269 & 419 \\
\hline & Total & 1650 & 53,687 & 55,337 \\
\hline 
\end{tabular}
\end{table}

\subsection{Experimental Setup}
We undertook two sets of experiments to analyse the performance of the proposed LiDAR-guided cross-attention band selection method. The first set of experiments compares our method with traditional hyperspectral band-selection methods, which do not consider the LiDAR data in the band-selection. The results are presented in terms of classification performance after fusion of the selected hyperspectral bands with the LiDAR data. The second set of experiments compares the results of our method with several hyperspectral and LiDAR fusion models without band selection. 

In the experiment, both Support Vector Machine (SVM) and Convolutional Neural Networks (CNN) were used for the classification task. The standard Libsvm toolbox was adopted for the SVM. For the CNN, a conventional 6-layer network was selected, with kernel size $3\times3$, dropout 0.4, a dense layer, and a ReLU activation function. A softmax function was used for classification purposes. The batch size was set to 32, the learning rate was 0.0001, and the epoch was 50. The classification performance was evaluated using three commonly used criteria, i.e., overall accuracy (OA), average accuracy (AA), and kappa coefficient (Kappa). 

\subsection{Implementation Details} 
The experiments were carried out on the PyTorch platform, and the model was trained on the GPU-enabled machine of the Colab Premium version to accelerate the computation.

On the Houston and Trento dataset, the size of the HSI sample patch was $9\times9$, Following the patch embedding method of the visual transformer, the sample size $9\times9\times144$ is divided into 144 pieces of $9\times9$ patches, and the kernel size is set to $1\times1$. with the number of patches equal to the number of HSI bands, ensuring a one-to-one correspondence between the patches and the spectral bands for effective band selection. LiDAR data were processed similarly with the patch size set to $9\times9$, and the patch number was 1.  For the MUUFL dataset, the patch size was reduced to $3\times3$. A linear projection layer transformed each patch into a flattened embedded representation. The embedding size was 256, ensuring that each spectral band was treated as a distinct entity. Consequently, both the HSI and LiDAR embedding were added with positional embedding of size 256 to retain spatial information.

Each self-attention module consisted of 3 layers with 8 heads, where the dimensionality of each head was 128 and the multilayer perceptron (MLP) dimension within the transformer blocks was 256. A cross-attention mechanism then integrated the information from both modalities, allowing each LiDAR patch to attend to all HSI patches, using the attention scores for band selection. Adam optimiser with a learning rate of 0.0001 was used to train the model. The model was trained with a batch size of 32.

Training was terminated after 50 epochs with an early stopping criterion to prevent overfitting. A cross-entropy loss function was employed to calculate the difference between the predicted class probabilities and the ground-truth labels. We report the results with and without augmented training data. In the former case,  the augmented samples were generated through transformations including 45- and 90-degree rotations and both vertical and horizontal flips. 

After training, the network cross-attention module assigned significance to each hyperspectral band based on its relevance to the corresponding LiDAR data. The cross-attention module outputs the attention weights, showing the relevance between the HSI and LiDAR data. The weights were then used to predict important band selection. Then, this subset of selected bands is used for subsequent data fusion tasks. 

\subsection{{Ablation Study}} 
{In this study, we investigate the incremental performance enhancements offered by our proposed cross-attention band selection model, particularly focusing on the accuracy of classification.
This investigation is achieved by sequentially omitting the cross-attention component and
instead employing alternatives such as hyperspectral only self-attention and a mixed self-attention mechanism that incorporates both LiDAR and hyperspectral imagery (HSI) within
the network architecture. A series of ablation experiments were conducted using the SVM classifier in the Houston 2013 data set to rigorously assess the contribution of the cross-attention module to the HSI and HSI + LiDAR classification tasks}.

{More specifically, the ablation study explores two different configurations of self-attention: Self-attention A leverages a composite importance score that amalgamates information from both HSI and LiDAR data for band selection. Conversely, Self-attention B uses only hyperspectral image data, where the selection of bands is guided by the importance scores derived from hyperspectral image self-attention post-model training. These methodologies were integrated into our model's framework, and were evaluated across both raw and augmented data scenarios. }

{Subsequent SVM classification tests were conducted to compare the efficacy of different band selection strategies, incorporating the varied band indices generated by the four examined methods.} 

{The results of these experiments are systematically presented in Table~\ref{uh2013ab}. We can see that the augmented data-enhanced cross-attention module secures the top performance ranking. This is closely followed by the model employing the cross-attention module trained on raw data. In contrast, the models utilizing Self-attention A and Self-attention B configurations exhibit comparatively diminished performance, underscoring the superior effectiveness of the cross-attention mechanism in our band selection model. }


\begin{table*}[htp!]
\centering
\caption{Ablation Analysis of the Proposed LiDAR Guided Band Selection Model With a Combination of Two Self Attention Modules On Houston 2013 Dataset .The best results of SVM are highlighted using \textcolor{blue}{blue color} and the second best results are highlighted using \textcolor{red}{red color.} }
\label{uh2013ab}
\begin{tabular}{|c|c|c|c|c|c|c|}
\hline
\textbf{Band Number}  & Metrics & Self-Attention A & Self-Attention B & Self-Attention+Fusion& Self-Attention+Fusion\\ 
& &Trained On HSI+LIDAR & Trained on HSI &OurMethod(Raw Data)& OurMethod(Augmented Data) \\
\hline
\hline
\multirow{3}{*}{30} &OA&0.9125&	0.9073&	\textcolor{red}{0.9201}	&\textcolor{blue}{0.9242}\\
 &AA& 0.9248	&0.9202&	\textcolor{red}{0.9301}&	\textcolor{blue}{0.9373} \\
&kappa& 0.9050&	0.8994&	\textcolor{red}{0.9133}&	\textcolor{blue}{0.9177}\\
\hline
\multirow{3}{*}{25} &OA&0.9125&	0.9033&	\textcolor{red}{0.9193}&	\textcolor{blue}{0.9266}\\
 &AA& {0.9248}&	0.9171&\textcolor{red}	{0.9284}&	\textcolor{blue}{0.9392}\\
&kappa&0.9050&	0.8950&	\textcolor{red}{0.9124}&	\textcolor{blue}{0.9203}\\
\hline
\multirow{3}{*}{20} &OA&0.9118&	0.9101&	\textcolor{red}{0.9198}&	\textcolor{blue}{0.9253}\\
 &AA& 0.9235&	0.9223&	\textcolor{red}{0.9283}&	\textcolor{blue}{0.9378} \\
&kappa& 0.9042&	0.9024&	\textcolor{red}{0.9124}&	\textcolor{blue}{0.9189}\\
\hline
\multirow{3}{*}{15} &OA&\textcolor{red}{0.9134} & 0.9064&  0.9019&\textcolor{blue}{0.9251}\\
 &AA& \textcolor{red}{ 0.9246} &{ 0.9242} &0.9187& \textcolor{blue}{0.9378} \\
&kappa& \textcolor{red}{0.9060} & 0.8984 &0.8935&\textcolor{blue}{ 0.9186}\\
\hline
\multirow{3}{*}{10} & OA & 0.8742 & \textcolor{red}{0.9097} &0.8938&\textcolor{blue}{ 0.9317} \\
  & AA & 0.8909 & \textcolor{red}{0.9209} &0.9107& \textcolor{blue}{0.9427} \\
  & Kappa & 0.8634 & \textcolor{red}{0.9019} &0.8848&\textcolor{blue}{ 0.9259} \\
\hline
\multirow{3}{*}{5} & OA & 0.8766 & 0.8762 &\textcolor{red}{0.9079}& \textcolor{blue}{0.9202} \\
  & AA & 0.8929& 0.8879 &\textcolor{red}{0.9210}& \textcolor{blue}{0.9335} \\
  & Kappa & 0.8660 & 0.8557 &\textcolor{red}{0.9000}& \textcolor{blue}{0.9134} \\
\hline

\end{tabular}\\
\end{table*}

{This ablation study conclusively affirms the pivotal role of the cross-attention module in optimising band selection for HSI and HSI + LiDAR fusion applications, highlighting its significant impact on enhancing classification accuracy.}

\subsection{Comparison with HSI band selection methods}
In this set of experiments, we fused bands selected by several recent band selection approaches and concatenated them with the corresponding LiDAR data before calling the classifiers. The compared band selection methods included low-rank representation-based band selection ($LRR-BS$)~\cite{yu2021semisupervised}, adaptive spatial spectral patch selection (ASPS)~\cite{wang2019hyperspectral}, orthogonal projection band selection (OPBS)~\cite{8320544}, optimal clustering framework (TRC-OC-FDPC)~\cite{wang2018optimal} and band selection network (BSNET)~\cite{cai2019bs}. We tested the performance of each approach with 1 to 50 selected bands and all bands used based on both SVm and 1DCNN.


\begin{figure*}[t]
\centering
\subfloat[]{\includegraphics[width=9cm, height=5.5cm]{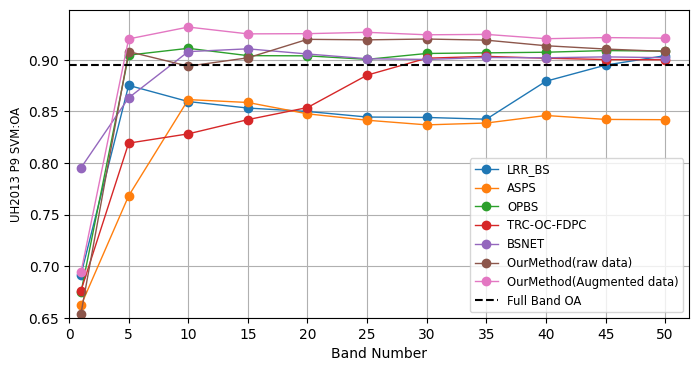}%
\label{fig_uh_svm_oa}}
\hfil
\subfloat[]{\includegraphics[width=9cm, height=5.5cm]{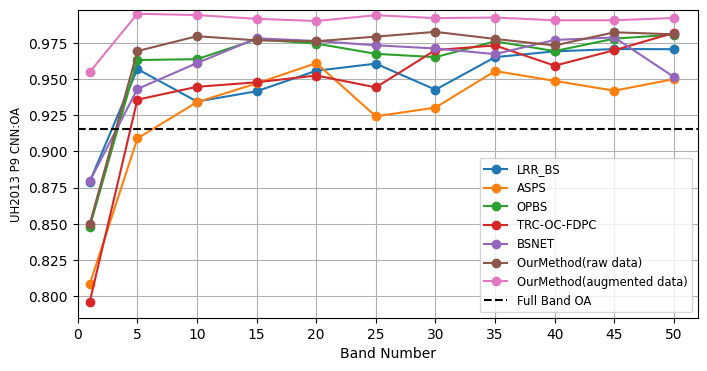}%
\label{fig_uh_cnn_oa}}
\caption{Classification results on the Houston 2013 data set after fusing selected bands with LiDAR data. (a) OA with SVM; (b) OA with CNN. }
\end{figure*}

\begin{table*}[htp!]
\centering
\caption{Classification performance on the Houston 2013 data set (HSI+LidAR) with the SVM classifier. The best results are highlighted using \textcolor{blue}{blue color} and the second best results are highlighted using \textcolor{red}{red color.} }
\label{uh2013svmcom}
\begin{tabular}{|c||c||c||c||c||c||c||c|c|}
\hline
\textbf{Metrics}  & \textbf{LRR\_BS} & 
\textbf{ASPS} & \textbf{OPBS} & 
\textbf{TRC-OC-FDPC} & \textbf{BSNET}  &
\textbf{Our Method}&\textbf{ Our Method} \\ 
& \cite{yu2021semisupervised} & \cite{wang2019hyperspectral} & \cite{8320544}& \cite{wang2018optimal} & \cite{cai2019bs} & \textbf{(Raw Data)}& \textbf{(Augmented Data)} \\
\hline
\hline
OA (30 bands)   & 0.8441&	0.8369&	0.9061&	0.9015&	0.9003&\textcolor{red}{0.9201}&\textcolor{blue}{0.9242} \\
AA (30 bands)   &0.8684&	0.8584&	0.9199&	0.9152&	0.9152&\textcolor{red}{0.9301}&	\textcolor{blue}{0.9373}\\
Kappa (30 bands)  & 0.8308&	0.8232&	0.8981&	0.8931&	0.8918&\textcolor{red}{0.9133}&	\textcolor{blue}{0.9177}\\
\hline

OA (25 bands) &0.8445&	0.8414&	0.9004&	0.8851&	0.9010&	\textcolor{red}{0.9193}&	\textcolor{blue}{0.9266}\\
AA (25 bands) &0.8695&	0.8624&	0.9155&	0.9002&	0.9157&	\textcolor{red}{0.9284}&	\textcolor{blue}{0.9392}\\
Kappa (25 bands) &0.8313&	0.8281&	0.8919&	0.8752&	0.8925&	\textcolor{red}{0.9124}&	\textcolor{blue}{0.9203}\\
\hline
OA (20 bands)&0.8499&	0.8476&	0.9038&	0.8537&	0.9056&\textcolor{red}{0.9198}&\textcolor{blue}{0.9253}\\
AA (20 bands)&0.8736&	0.7675&	0.9173	&0.8726&	0.9193&	\textcolor{red}{0.9283}&\textcolor{blue}{0.9378}\\
Kappa (20 bands)&0.8371	&0.8348	&0.8955&	0.8412&	0.8976&	\textcolor{red}{0.9124}&\textcolor{blue}{0.9189}\\
\hline
OA (15 bands) &0.8532&	0.8586&	0.9040&	0.8420&	0.9106&	\textcolor{red}{0.9019}&\textcolor{blue}{0.9251}\\
AA (15 bands)  &0.8770&	0.8763&	0.9175&	0.8628&	0.9231&	\textcolor{red}{0.9167}&\textcolor{blue}{0.9378}\\
Kappa (15 bands)&0.8408&	0.8466&	0.8958&	0.8286&	0.9030&	\textcolor{red}{0.8935}&\textcolor{blue}{0.9186}\\
\hline
OA (10 bands) & 0.8595&	0.8614&	\textcolor{red}	{0.9111}&	0.8282&	0.9079&	0.8938&	\textcolor{blue}{0.9317} \\
AA (10 bands) &0.8822&	0.8776&	0.9233&	0.8515&		\textcolor{red}{0.9203}&	0.9107&	\textcolor{blue}{0.9427} \\
Kappa (10 bands) &0.8476&	0.8496&	0.9035&	0.8138&		\textcolor{red}{0.9000}&	0.8848&	\textcolor{blue}{0.9259} \\
\hline
OA (5 bands)  & 0.8752&	0.7681&		\textcolor{red}0.9045&	0.8192&	0.8631&	0.9079&	\textcolor{blue}{0.9202} \\
AA (5 bands) & 0.8934&	0.7994&	0.9171&	0.8424&	0.8819&		\textcolor{red}{0.9210}&	\textcolor{blue}{0.9335} \\
Kappa (5 bands) &0.8646&	0.7492&	0.8963&	0.8041&	0.8515&		\textcolor{red}{0.9000}&	\textcolor{blue}{0.9134} \\
\hline
OA (1 band) & 0.6918&	0.6619&	0.6745&	0.6762&	\textcolor{blue}{0.7952}&0.6539 &\textcolor{red}{0.6939}\\
AA (1 band) &\textcolor{blue}{0.7361}&	0.7014&	0.7214	&0.7221&	0.8237&0.6983 &	\textcolor{red}{0.7242}\\
Kappa (1 band) & 0.6677&	0.6355&	0.6495&	0.6510&	\textcolor{blue}{0.7785}&0.6269 &	\textcolor{red}{0.6693}\\
\hline
\end{tabular}\\
\end{table*}

\begin{table*}[htp!]
\centering
\caption{Classification performance on the Houston 2013 data set (HSI+LidAR) with the CNN classifier. The best results are highlighted using \textcolor{blue}{blue color} and the second best results are highlighted using \textcolor{red}{red color.}}
\label{uh2013cnncom}
\begin{tabular}{|c||c||c||c||c||c||c||c|c|}
\hline
\textbf{Metrics}  & \textbf{LRR\_BS} & 
\textbf{ASPS} & \textbf{OPBS} & 
\textbf{TRC-OC-FDPC} & \textbf{BSNET}  &
\textbf{Our Method}&\textbf{ Our Method} \\ 
& \cite{yu2021semisupervised} & \cite{wang2019hyperspectral} & \cite{8320544} & \cite{wang2018optimal} & \cite{cai2019bs} & \textbf{(Raw Data)} & \textbf{(Augmented Data)} \\
\hline
\hline
OA (30 bands)&0.9427&	0.9303&	0.9652&	0.9701&	0.9712&	\textcolor{red}{0.9827}&\textcolor{blue}{0.9922}\\
AA (30 bands)&0.9521&	0.9384&	0.9701&	0.9704&	0.9687&	\textcolor{red}{0.9815}&\textcolor{blue}{0.9940}\\
Kappa (30 bands)&0.9379&0.9244&	0.9622&	0.9676&	0.9688&	\textcolor{red}{0.9814}&\textcolor{blue}{0.9915}\\
\hline
OA (25 bands)&0.9607&	0.9243&	0.9675&	0.9443&	0.9734&	\textcolor{red}{0.9794}&\textcolor{blue}{0.9942}\\
AA (25 bands) &0.9647&	0.9292&	0.9696&	0.9515&	0.9763&	\textcolor{red}{0.9794}&\textcolor{blue}{0.9955}\\
Kappa (25 bands)&0.9574&0.9179&	0.9647&	0.9395&	0.9711&	\textcolor{red}{0.9777}&\textcolor{blue}{0.9938}\\
\hline
OA (20 bands) &0.9558&	0.9611&	0.9746&	0.9525&	\textcolor{red}{0.9764}&0.9762&\textcolor{blue}{0.9902}\\
AA (20 bands) &0.9612&	0.9655&	0.9732&	0.9548&	\textcolor{red}{0.9755}&{0.9786}&\textcolor{blue}{0.9904}\\
Kappa (20 bands)&0.9520&	0.9577&	0.9724&0.9484&\textcolor{red}{0.9744}&	0.9742&	\textcolor{blue}{0.9894}\\
\hline
OA (15 bands) &0.9416&0.9470&0.9775&	0.9478&	\textcolor{red}{0.9782}&0.9768&\textcolor{blue}{0.9917}\\
AA (15 bands)&0.9513&0.9530&0.9690	&0.9537&\textcolor{red}{0.9789}&0.9773&\textcolor{blue}{0.9935}\\
Kappa (15 bands)&0.9366&0.9425&	0.9756&	0.9433&	\textcolor{red}{0.9763}&0.9748&\textcolor{blue}{0.9910}\\
\hline
OA (10 bands) & 0.9345&0.9341&	0.9638&	0.9447&	0.9611&	\textcolor{red}{0.9797}&\textcolor{blue}{0.9943} \\
AA (10 bands) & 0.9440&	0.9418&	0.9670&	0.9489&	0.9646&	\textcolor{red}{0.9795}&\textcolor{blue}{0.9956}\\
Kappa (10 bands) & 0.9289&	0.9284&	0.9607&	0.9399&	0.9577&	\textcolor{red}{0.9780}&\textcolor{blue}{0.9939} \\
\hline 
OA (5 bands) & 0.9570&	0.9090&	0.9632&	0.9358&	0.9432&	\textcolor{red}{0.9694}&\textcolor{blue}{0.9952} \\
AA (5 bands) & 0.9583&	0.9157&	0.9614&	0.9450&	0.9512&	\textcolor{red}{0.9735}&\textcolor{blue}{0.9963}\\
Kappa (5 bands) & 0.9533&	0.9012&	0.9600&	0.9303&	0.9383&	\textcolor{red}{0.9668}&\textcolor{blue}{0.9948}\\
\hline 
OA (1 band) & 0.8791&0.8081&0.8477&0.7959&	\textcolor{red}{0.8794}&0.8496&\textcolor{blue}{0.9549} \\
AA (1 band)  &0.8905&	0.8352&	0.8670&	0.8229&	\textcolor{red}{0.8965}&0.8655&\textcolor{blue}{0.9639} \\
Kappa (1 band) & 0.8688&	0.7919&	0.8349&	0.7790&	\textcolor{red}{0.8692}&0.8371&\textcolor{blue}{0.9510}   \\
\hline 
\hline
\end{tabular}\\
\end{table*}

\subsubsection{\textbf{Results on the Houston 2013 data set}}
We show the experimental results in Fig.~\ref{fig_uh_svm_oa} for OA and in Tables~\ref{uh2013svmcom} 
and~\ref{uh2013cnncom} for OA, AA and Kappa. Please note that due to space limitations, the curves on AA and
kappa are not reported here, but can be accessed on GitHub\footnote{\url
{https://github.com/Judyxyang/LiDAR-Guided-Band-Selection/tree/main/img}}.
Our methodology has consistently outperformed other band selection techniques in evaluations with SVM and CNN classifiers, demonstrating its robustness and effectiveness. When using models trained on raw data, our approach not only maintains stable performance but also achieves superior accuracy and Kappa scores, particularly when selecting from 15 to 40 bands. In contexts where CNN classifiers are employed, although BSNET shows promising initial performance, our method, applied to raw data, exceeds it in efficiency with an increasing number of bands, specifically at 5, 15,20, 25, 30, and 35 bands. The performance enhancement becomes even more pronounced with models trained on augmented data. In such scenarios, both the SVM and CNN classifiers exhibit exceptional performance, significantly outperforming competing methods in nearly all band selection scenarios. This empirical evidence strongly supports our hypothesis that band selection, when combined with LiDAR data, significantly increases the effectiveness of the classification process after multimodal fusion.


\begin{figure*}[htp!]
\centering
\subfloat[]{\includegraphics[width=9cm, height=5.5cm]{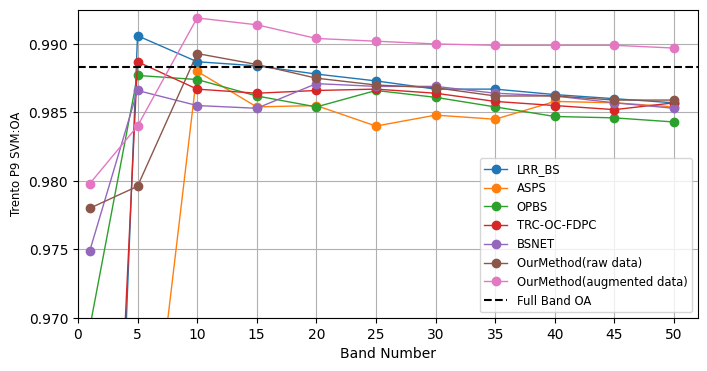}%
\label{fig_tr_svm_oa}}
\hfil
\subfloat[]{\includegraphics[width=9cm, height=5.5cm]{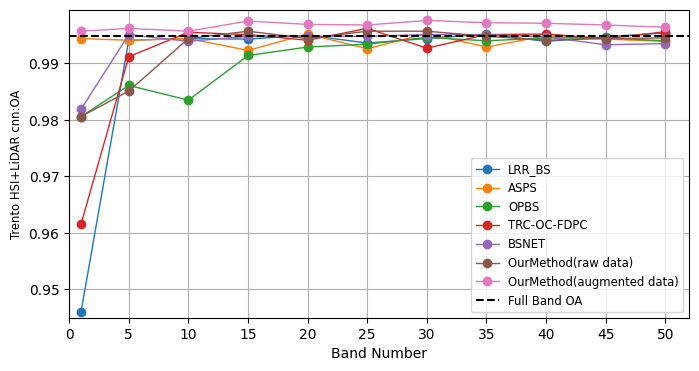}%
\label{fig_tr_cnn_oa}}
\caption{Classification results on the Trento data set after fusing selected bands with LiDAR data. (a) OA with SVM; (b) OA with CNN}
\label{fig_tr_oa}
\end{figure*}

\begin{table*}[htp!]
\centering
\caption{Classification performance on the Trento data set (HSI+LidAR) with the SVM classifier. The best results are highlighted using \textcolor{blue}{blue color} and the second best results are highlighted using \textcolor{red}{red color.}}
\label{tr-svm-com}
\begin{tabular}{|c||c||c||c||c||c||c||c|c|}
\hline
\hline
\textbf{Metrics}  & \textbf{LRR\_BS} & 
\textbf{ASPS} & \textbf{OPBS} & 
\textbf{TRC-OC-FDPC} & \textbf{BSNET}  &
\textbf{Our Method}&\textbf{Our Method} \\ 
&\cite{yu2021semisupervised}& \cite{wang2019hyperspectral} &\cite{8320544} &\cite{wang2018optimal} & \cite{cai2019bs} & \textbf{(Raw Data)} & \textbf{(Augmented Data)} \\

\hline
 OA (30 bands)&0.9867&	0.9848&	0.9861&	0.9864&	\textcolor{red}{0.9869}&	0.9868&	\textcolor{blue}{0.9900}\\
AA (30 bands)&0.9786&	0.9768&	0.9791&	0.9793&	\textcolor{red}{0.9798}&	0.9793&	\textcolor{blue}{0.9862}\\
Kappa(30 bands)&0.9821&	0.9795&	0.9813&	0.9818&	\textcolor{red}{0.9824}&	0.9823&	\textcolor{blue}{0.9866}\\
\hline
 OA (25 bands)&\textcolor{red}{0.9873}&	0.9840&	0.9866&	0.9867&	0.9869&	0.9870&	\textcolor{blue}{0.9902}\\
AA (25 bands)&\textcolor{red}{0.9798}&	0.9758&	0.9802&	0.9790&	0.9798	&0.9795&	\textcolor{blue}{0.9866}\\
Kappa(25 bands)&0.9819&	0.9785&	0.9820&	0.9822&	0.9824&	\textcolor{red}{0.9826}&	\textcolor{blue}{0.9869}\\
\hline
 OA (20 bands)&\textcolor{red}{0.9878}&	0.9855&	0.9854&	0.9866&	0.9871&	0.9875&	\textcolor{blue}{0.9904}\\
AA (20 bands)&0.9819&	0.9783&	0.9797&	0.9779&	0.9802&	\textcolor{red}{0.9804}&	\textcolor{blue}{0.9870}\\
Kappa(20 bands)&\textcolor{red}{0.9836}&	0.9806&	0.9804&	0.9821&	0.9828&	0.9833&\textcolor{blue}{0.9871}\\
\hline
 OA (15 bands)&0.9884&	0.9854&	0.9862&	\textcolor{red}{0.9864}&	0.9853&	0.9885&	\textcolor{blue}{0.9914}\\
AA (15 bands)&0.9817&	0.9784&	0.9806&	0.9783&	0.9798&	\textcolor{red}{0.9817}&	\textcolor{blue}{0.9881}\\
Kappa(15 bands)&0.9844&	0.9804&	0.9815&	0.9817&	0.9803&	\textcolor{red}{0.9845}&	\textcolor{blue}{0.9885}\\
\hline
 OA (10 bands) &0.9887&	0.9813&	0.9874&	0.9804&	0.9855&	\textcolor{red}{0.9893}&	\textcolor{blue}{0.9919} \\
AA (10 bands)  & 0.9805&	0.9726&	0.9816&	0.9716&	0.9815&	\textcolor{red}{0.9849}&	\textcolor{blue}{0.9886}\\
Kappa(10 bands)  & 0.9848&	0.9748&	0.9831&	0.9737&	0.9805	&\textcolor{red}{0.9856}&	\textcolor{blue}{0.9888}\\
\hline
 OA (5 bands)  & 0.9806&	0.9512&	0.9877&	\textcolor{blue}{0.9887}&	\textcolor{red}{0.9866}&	0.9796&	0.9840 \\
 AA (5 bands)  & \textcolor{blue}{0.9845}&	0.9316&	0.9812&	0.9816&	\textcolor{red}{0.9817}&	0.9713&	0.9778 \\
 Kappa (5 bands)  &\textcolor{blue}{ 0.9874}&	0.9339&	0.9834&	0.9848&	0.9820	&0.9726&	\textcolor{red}{0.9786 }\\
\hline
 OA (1 band)  & 0.9018&	0.9030&	0.9693&	0.9150	&0.9749&	\textcolor{red}{0.9780}&	\textcolor{blue}{0.9798} \\
AA (1 band) & 0.7753&	0.8946&	0.9565&	0.9266&	0.9631&	\textcolor{red}{0.9657}&	\textcolor{blue}{0.9698}\\
Kappa (1 band)  &0.8677&	0.8712&\textcolor{red}{0.9587}&	0.8873&	0.9663&	0.9704&	\textcolor{blue}{0.9729}\\
\hline
\hline
\end{tabular}\\
\end{table*}

\begin{table*}[htp!]
\centering
\caption{Classification performance on the Trento data set (HSI+LidAR) with the CNN. The best results are highlighted using \textcolor{blue}{blue color} and the second best results are highlighted using \textcolor{red}{red color.}}
\label{tr-cnn-com}
\begin{tabular}{|c||c||c||c||c||c||c||c|c|}
\hline
\hline
\textbf{Metrics} & \textbf{LRR\_BS} & 
\textbf{ASPS} & \textbf{OPBS} & 
\textbf{TRC-OC-FDPC} & \textbf{BSNET}  &
\textbf{Our Method}&\textbf{Our Method} \\ 
& \cite{yu2021semisupervised} &\cite{wang2019hyperspectral} &\cite{8320544} &\cite{wang2018optimal} &\cite{cai2019bs} & \textbf{(Raw Data)}& \textbf{(Augmented Data)} \\
\hline
OA (30 bands)&0.9944& 	0.9953& 	0.9946& 	0.9927& 	0.9951& 	\textcolor{red}{0.9957}&	\textcolor{blue}{0.9976}\\
AA (30 bands)&0.9915& 	\textcolor{red}{0.9930}& 	0.9911& 	0.9872& 	0.9925& 	0.9927&	\textcolor{blue}{0.9962}\\
Kappa (30 bands)&0.9926& 	0.9938& 	0.9928	& 0.9902& 	0.9934& 	\textcolor{red}{0.9943}&	\textcolor{blue}{0.9969}\\
\hline
OA (25 bands)&0.9936& 	0.9926& 	0.9934& 	\textcolor{red}{0.9962}& 	0.9948& 	0.9957&	\textcolor{blue}{0.9968}\\
AA (25 bands 25)&0.9906& 	0.9904& 	0.9898& 	0.9908& 	0.9904	& \textcolor{red}{0.9929}&	\textcolor{blue}{0.9948}\\
Kappa (25 bands)&0.9915& 	0.9902	& 0.9911& 	\textcolor{red}{0.9949}	& 0.9931	& 0.9943&\textcolor{blue}{0.9958}\\
\hline
OA (20 bands)&0.9950& 	\textcolor{red}{0.9952}& 	0.9929& 	0.9941& 	0.9948& 	0.9944&	\textcolor{blue}{0.9969}\\
AA (20 bands)&0.9921& 	\textcolor{red}{0.9925}& 	0.9884& 	0.9914& 	0.9923& 	0.9914&	\textcolor{blue}{0.9950}\\
Kappa (20 bands)&\textcolor{red}{0.9933}& 	0.9936& 	0.9905& 	0.9922& 	0.9930& 	0.9925&\textcolor{blue}{0.9959}\\
\hline
OA (15 bands)&0.9943& 	0.9923& 	0.9914& 	0.9951& 	0.9948& 	\textcolor{red}{0.9957}&	\textcolor{blue}{0.9975}\\
AA (15 bands)&0.9922& 	0.9899& 	0.9867& 	0.9846& 	0.9922& 	\textcolor{red}{0.9931}&	\textcolor{blue}{0.9960}\\
Kappa (15 bands)&0.9923& 	0.9898& 	0.9884& 	0.9935& 	0.9931& 	\textcolor{red}{0.9943}&	\textcolor{blue}{0.9967}\\
\hline
OA (10 bands)  & 0.9944& 	0.9944& 	0.9835& 	0.9955& 	0.9939& 	\textcolor{red}{0.9944} &\textcolor{blue}{0.9957}\\
AA (10 bands) & 0.9900& 	0.9918& 	0.9754& 	0.9901& 	\textcolor{red}{0.9907}& 	0.9915&\textcolor{blue}{0.9934}\\
Kappa (10 bands) & 0.9925& 	0.9926& 	0.9778& 	0.9940& 	0.9919& 	0.9927&	\textcolor{blue}{0.9943} \\
\hline
OA (5 bands)  &0.9941& 	0.9941& 0.9861& 0.9912& \textcolor{red}{0.9951}& 0.9851 &\textcolor{blue}{0.9962} \\
AA (5 bands)  & 0.9914& 0.9922& 0.9862& 0.9909& \textcolor{red}{0.9922}& 0.9736&	\textcolor{blue}{0.9964}\\
Kappa (5 bands)  &0.9921& 0.9922& 0.9813& 0.9885& \textcolor{red}{0.9934}& 0.9800&\textcolor{blue}{0.9950}\\
\hline
OA (1 band)  & 0.9461& 	\textcolor{red}{0.9944}& 0.9806& 0.9616& 0.9819& 0.9806&\textcolor{blue}{0.9957} \\
AA (1 band)  & 0.9007& 	\textcolor{red}{0.9911}& 0.9710& 0.9416& 0.9694& 0.9659& \textcolor{blue}{0.9946}\\
Kappa (1 band) & 0.9278& \textcolor{red}{0.9929}& 0.9740& 0.9486& 0.9757& 0.9740&\textcolor{blue}{0.9943}\\
\hline
\hline
\end{tabular}\\
\end{table*}

\subsubsection{\textbf{Results on the Trento data set}}
Fig.~\ref{fig_tr_oa} and Tables~\ref{tr-svm-com} and~\ref{tr-cnn-com} show OA, AA, and Kappa with various bands selected for the SVM and the CNN classifiers. According to the results, our method with the raw training dataset outperforms other methodologies using the SVM classifier when the band numbers are 1, 5 and 10 in most criteria. The results demonstrate a remarkable capability to discern pertinent features from a constrained band set, a crucial attribute in contexts with limited data availability or computational resources. When trained with augmented data, our method performs the best in all cases and presents its exceptional classification capability. This implies that the attention-based mechanism can work effectively with fused hyperspectral bands and LiDAR data. In particular, CNN achieved close to perfect classification results, validating the strength of a deep learning-based approach that integrates attention mechanisms with deep neural networks.

\subsubsection{\textbf{Results on the MUUFL data set }}
\begin{figure*}[!htp!]
\centering
\subfloat[]{\includegraphics[width=9cm, height=5cm]{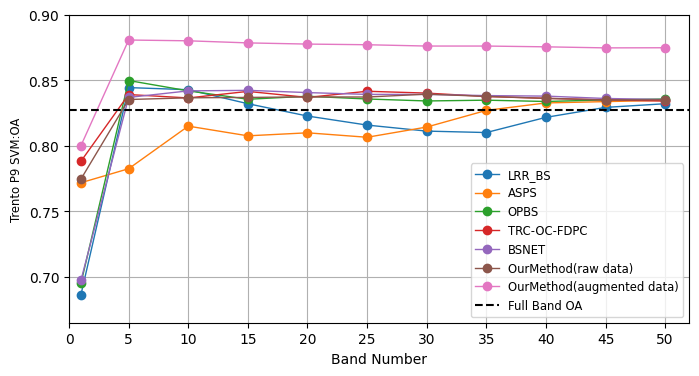}
\label{fig_mf_svm_oa}}
\hfil
\subfloat[]{\includegraphics[width=9cm, height=5cm]
{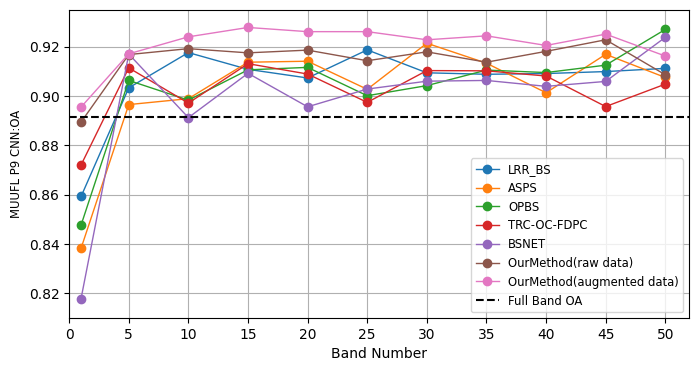}%
\label{fig_mf_cnn_oa}}
\caption{Classification results on the MUUFL data set after fusing selected bands with LiDAR data. (a) OA with SVM; (b) OA with CNN.}
\label{fig_mf_oa}
\end{figure*}

The results on the MUUFL dataset are shown in Fig.~\ref{fig_mf_oa}, Table~\ref{mf-svm-com} and Table~\ref{mf-cnn-com}. The evaluation of the SVM classifier based on the MUUFL data set demonstrates that our method with limited training data exhibits close performance to the top performing methods. Meanwhile, when trained on the augmented data, our method consistently exceeds alternative techniques such as ASPS, OPBS, and BSNET across various band counts and with both SVM and CNN classifiers. 

The experiments described above show that the use of LiDAR to guide band selection is a key factor in our method. LiDAR provides additional spatial and depth information that can enhance the feature selection process, leading to more accurate and reliable classifications. The implementation of a cross-attention mechanism contributes to the method's effectiveness. This mechanism can better capture the relationships between different types of data (HSI and LiDAR in this case), leading to more informative feature selection and, consequently, better classification performance.

\begin{table*}[htp!]
\centering
\caption{Classification performance on the MUUFL data set (HSI+LidAR) with the SVM classifier. The best results are highlighted using \textcolor{blue}{blue color} and the second best results are highlighted using \textcolor{red}{red color.}}
\label{mf-svm-com}
\begin{tabular}{|c||c||c||c||c||c||c||c|c|}
\hline
\textbf{Metrics}  & \textbf{LRR\_BS} & 
\textbf{ASPS} & \textbf{OPBS} & 
\textbf{TRC-OC-FDPC} & \textbf{BSNET}  &
\textbf{Our Method}&\textbf{Our Method} \\ 
& \cite{yu2021semisupervised}&\cite{wang2019hyperspectral} &\cite{8320544} &\cite{wang2018optimal} & \cite{cai2019bs} & \textbf{(Raw Data)} & \textbf{(Augmented Data)} \\
\hline
\hline
OA (30 bands)&0.8112&	0.8143&	0.8342&	0.8402&	0.8392&\textcolor{red}{	0.8394}&	\textcolor{blue}{0.8761}\\
AA (30 bands)&0.8522&	0.8534&	0.8649&	0.8752&	0.8732&	\textcolor{red}{0.8745}&\textcolor{blue}{	0.8993}\\
Kappa (30 bands)& 0.7603&	0.7639&	0.7885&	\textcolor{red}{0.7960}&	0.7948	&0.7949&	\textcolor{blue}{0.8398}\\
\hline
OA (25 bands)&0.8158&	0.8065&	0.8357&\textcolor{red}{	0.8416}&	0.8392&	0.8372&	\textcolor{blue}{0.8771}\\
AA ((25 bands)&0.8539&	0.8485&	0.8664&	\textcolor{red}{0.8766}&	0.8721	&0.8717	&\textcolor{blue}{0.8994}\\
Kappa ((25 bands5)& 0.7660&	0.7543&\textcolor{red}{	0.7905}&0.7977	&0.7945&	0.7922&\textcolor{blue}{	0.8411}\\
\hline
OA (20 bands)&0.8227&	0.8099&	0.8378&	0.8369&	\textcolor{red}{0.8406}&	0.8373&	\textcolor{blue}{0.8776}\\
AA (20 bands)&0.8615&	0.8503&	0.8688&	\textcolor{red}{0.8750}&	0.8716&	0.8726&	\textcolor{blue}{0.8980}\\
Kappa (20 bands)&0.7747&	0.7585&	0.7930&	0.7922&	\textcolor{red}{0.7965}&	0.7923&\textcolor{blue}{	0.8417}\\
\hline
OA (15 bands)&0.8321&	0.8076&	0.8355&	0.8415&	\textcolor{red}{0.8423}&	0.8369&\textcolor{blue}{	0.8785}\\
AA (15 bands)&0.8717&	0.8497&	0.8679&	\textcolor{red}{0.8773}&0.8721&	0.8735&\textcolor{blue}{	0.8991}\\
Kappa (15 bands)& 0.7864&	0.7558&	0.7903&	0.7978&	\textcolor{red}{0.7986}&	0.7919&\textcolor{blue}{	0.8428}\\
\hline
OA (10 bands) & \textcolor{red}{0.8428}&	0.8150&	0.8419&	0.8365&	0.8419&	0.8366&\textcolor{blue}{	0.8801}\\
AA (10 bands) & \textcolor{red}{0.8806}&	0.8554&	0.8755&	0.8728&	0.8737&	0.8729&	\textcolor{blue}{0.9002} \\
Kappa (10 bands)  &\textcolor{red}{0.7995}&	0.7648	&0.7983&0.7916&	0.7981&	0.7916&	\textcolor{blue}{0.8448}\\
\hline
OA (5 bands)  & 0.8444&	0.7825&\textcolor{red}{	0.8498}&0.8391&	0.8369&	0.8353&\textcolor{blue}{0.8807}\\
AA (5 bands)  & 0.8816&	0.8336&	\textcolor{red}{0.8856}&0.8758&	0.8702&	0.8716&	\textcolor{blue}{0.9007}\\
Kappa (5 bands)  &0.8013&	0.7254&	\textcolor{red}{0.8082}&0.7950&	0.7920&	0.7900&\textcolor{blue}{0.8455}\\
\hline
OA (1 band)  &0.6864&	0.7718&	0.6951&\textcolor{red}{	0.7882}&	0.6974&	0.7746&\textcolor{blue}{0.8001} \\
AA (1 band) & 0.7737&	0.8264&	0.7801&	\textcolor{blue}{0.8379}&	0.7799&	0.8204&	\textcolor{red}{0.8310}\\
Kappa (1 band)  &0.6129&	0.7119&	0.6229&\textcolor{red}{	0.7324}&	0.6254&	0.7152&	\textcolor{blue}{0.7443}\\
\hline
\hline
\end{tabular}\\
\end{table*}

\begin{table*}[htp!]
\centering
\caption{Classification performance on the MUUFL data set (HSI+LidAR) with the CNN classifier. The best results are highlighted using \textcolor{blue}{blue color} and the second best results are highlighted using \textcolor{red}{red color.}}
\label{mf-cnn-com}
\begin{tabular}{|c||c||c||c||c||c||c||c|c|}
\hline
\hline
\textbf{Metrics}  & \textbf{LRR\_BS} & 
\textbf{ASPS} & \textbf{OPBS} & 
\textbf{TRC-OC-FDPC} & \textbf{BSNET}  & 
\textbf{Our Method}&\textbf{ Our Method} \\ 
& \cite{yu2021semisupervised}&\cite{wang2019hyperspectral} &\cite{8320544} &\cite{wang2018optimal} & \cite{cai2019bs} & \textbf{(Raw Data)} & \textbf{(Augmented Train)} \\
\hline
OA (30 bands)&0.9094&\textcolor{red}{0.9214}&	0.9042&	0.9103&	0.9060&	0.9179&	\textcolor{blue}{0.9228}\\
AA (30 bands)& 0.8976&	0.9138&	0.8879&	0.8931&	0.8962&	\textcolor{blue}{0.9297}&	\textcolor{red}{0.9230}\\
Kappa (30 bands)& 0.8814&	\textcolor{red}{0.8970}&	0.8748&	0.8642&	0.8583&	0.8930&	\textcolor{blue}{0.8993}\\
\hline
OA (25 bands)&\textcolor{red}{0.9187}&	0.9027&	0.9001&	0.8974&	0.9028&	0.9143&	\textcolor{blue}{0.9261}\\
AA (25 bands)& 0.9030&	0.8973&	0.8925&	0.8796&	0.9051&	\textcolor{red}{0.9194}&\textcolor{blue}{	0.9346}\\
Kappa (25 bands)& \textcolor{red}{0.8935}&	0.8729&	0.8704&	0.8445&	0.8548&	0.8888&	\textcolor{blue}{0.9036}\\
\hline
OA (20 bands)&0.9073&	0.9141&	0.9116&	0.9089&	0.8956&	\textcolor{red}{0.9186}&	\textcolor{blue}{0.9261}\\
AA (20 bands)&0.9081&	0.9028&	0.9036&	0.9020&	0.9090&	\textcolor{red}{0.9304}&	\textcolor{blue}{0.9346}\\
Kappa (20 bands)& 0.8769&	0.8876&	0.8846&	0.8609&	0.8443&	\textcolor{red}{0.8936}&\textcolor{blue}{	0.9036}\\
\hline
OA (15 bands)& 0.9108&	0.9137&	0.9106&	0.9131&	0.9092	&\textcolor{red}{0.9175}&	\textcolor{blue}{0.9278}\\
AA (15 bands)& 0.9055&	0.8923&	0.9057&	0.8931&	0.9011&	\textcolor{red}{0.9238}&	\textcolor{blue}{0.9300}\\
Kappa (15 bands)& 0.8834&	0.8872&	0.8834&	0.8678&	0.8621&	\textcolor{red}{0.8924}&\textcolor{blue}{	0.9054}\\
\hline
OA (10 bands) & 0.9176&	0.8989&	0.8983&	0.8971&	0.8912&	\textcolor{red}{0.9192}&\textcolor{blue}{	0.9240} \\
AA (10 bands) &0.9121&	0.8875&	0.8912&	0.8970&	0.8828&	\textcolor{red}{0.9283}&\textcolor{blue}{	0.9306}\\
Kappa (10 bands) & 0.8926&	0.8679&	0.8673&	0.8449&	0.8364&	\textcolor{red}{0.8947}&	\textcolor{blue}{0.9006}\\
\hline
OA (5 bands) & 0.9033&	0.8965&	0.9063&	0.9114&	0.9170&	\textcolor{red}{0.9168}&	\textcolor{blue}{0.9171} \\
AA (5 bands) & 0.8956&	0.8933&	0.9051&	\textcolor{red}{0.9142}	&0.9057	&0.9115&	\textcolor{blue}{0.9266}\\
Kappa (5 bands) &0.8740&	0.8652&	0.8775&	0.8662&	0.8739&\textcolor{red}{	0.8909}	&\textcolor{blue}{0.8937}\\
\hline
OA (1 band) & 0.8592&	0.8382&	0.8476&	0.8720&	0.8178&	\textcolor{red}{0.8891}&	\textcolor{blue}{0.8955} \\
AA (1 band)   & 0.8618&	0.8224&	0.8566&	0.8842&	0.7727&	\textcolor{blue}{0.9009}&	\textcolor{red}{0.8839} \\
Kappa (1 band) &0.8164&	0.7863&	0.8015&	0.8099&	0.7277&	\textcolor{blue}{0.8562}&	\textcolor{red}{0.8432}\\
\hline
\hline
\end{tabular}\\
\end{table*}

{The overall MUUFLE classification performance is lower than those of the other two datasets. We believe the reason could be that the MUFFL dataset is more challenging than the other two datasets. This is also observed in other papers on HSI and LiDAR fusion.}

\subsection{Comparison with HSI and LiDAR Data Fusion Methods}
The fusion models selected for this comparative study include  the CNN-HSI model proposed by Yu et al. (2017)~\cite{yu2017convolutional} and further developed by Mohla et al. (2020)~\cite{mohla2020fusatnet} , the CoupledCNN model from Hang et al. (2020)~\cite{hang2020classification}, FuseAtNet as detailed in Mohla et al. (2020)~\cite{mohla2020fusatnet}, and a general multimodal deep learning (MDL-RS) framework including the early-fusion, middle-fusion and late-fusion models explored in Hong et al. (2020)~\cite{hong2020more}. Additional comparisons are made with EndNet (2020)~\cite{hong2020deep} and the collaborative convolutional learning model (CCL) presented by Jia et al. (2023)~\cite{jia2023collaborative}.

These state-of-the-art fusion models typically operate on full-band HSI hyperspectral image data combined with LiDAR. All experimental comparisons adhere to the same test settings as those delineated for the CCL configuration~\cite{jia2023collaborative}. Instead, our method uses only 10 HSI bands and fuses them with the LiDAR data.  For our model, band selection integrating 1DCNNCNN, and comparative models, analysis was carried out using a uniform patch size of 9~\cite{jia2023collaborative}. 

\subsubsection{\textbf{Results on the Houston 2013 data set}}
In the comprehensive analysis of the Houston 2013 dataset, our method has demonstrated exceptional performance, surpassing other fusion methods with a large margin. As highlighted in Table~\ref{uh2013_fusion}, our method has significantly outperformed alternative approaches that use all hyperspectral bands. In particular, our method trained on the augmented data has achieved an impressive Overall Accuracy (OA) of 0.9943, an Average Accuracy (AA) of 0.9956, and a Kappa coefficient of 0.9939. The results underscore the effectiveness and robustness of our method in describing hyperspectral images when using fewer selected bands.

\begin{table*}[htp!]
\centering
\caption{Classification performance on the Houston 2013 data set. All other bands used all bands in data fusion, while our method used only 10 selected bands. The best results are highlighted using \textcolor{blue}{blue color} and the second best results are highlighted using \textcolor{red}{red color.}}
\fontsize{9}{11}\selectfont
\begin{tabular}{|c||c||c||c||c||c||c||c||c|}
\hline 
\hline 
Metrics & CNN-HSI & CoupledCNN& Middle\_Fusion  & EndNet& FusAtNet & CCL & Our Method& Our Method\\
&\cite{yu2017convolutional}&\cite{hang2020classification} &\cite{hong2020more}&\cite{hong2020deep} &\cite{mohla2020fusatnet}&\cite{jia2023collaborative}&(Raw Data)&{(Augmented Data)} \\
\hline 
OA &0.8123 & 0.8835&0.8576 & 0.8098 & 0.7693 & 0.9215& \textcolor{red}{0.9797}&\textcolor{blue}{ 0.9943}\\

AA & 0.8309 & 0.8989 &0.8758 & 0.8166 & 0.7887 & 0.9306 & \textcolor{red}{0.9795}&\textcolor{blue}{0.9956} \\

kappa &0.7973 & 0.8741 & 0.8461 & 0.7944 & 0.7508 & 0.9151 & \textcolor{red}{0.9780}&\textcolor{blue}{0.9939}\\
\hline
\hline 
\end{tabular}
\label{uh2013_fusion}
\end{table*}

\subsubsection{\textbf{Results on the Trento data set}}
As shown in Table~\ref{trento_fusion}, the comparative assessment of classification performance on the Trento data set presents that our method with the raw training data has better performance than other methods in terms of AA, and are slighly worse on OA and Kappa. Meanwhile, when trained on the augmented dataset, our method achieved the highest classification accuracy. It leads the scoreboard in OA, AA and Kappa, demonstrating its superior capability to synthesise and classify hyperspectral data with a reduced set of 10 selected bands.

\begin{table*}[htp!]
\centering
\caption{Classification performance on the Trento data set. All other bands used all bands in data fusion, while our method used only 10 selected bands. The best results are highlighted using \textcolor{blue}{blue color} and the second best results are highlighted using \textcolor{red}{red color.}}
\fontsize{9}{11}\selectfont
\begin{tabular}{|c||c||c||c||c||c||c||c||c|}
\hline 
\hline 
Metrics & CNN-HSI & CoupledCNN& Middle\_Fusion  & EndNet& FusAtNet & CCL & Our Method& Our Method\\
&\cite{yu2017convolutional}&\cite{hang2020classification} &\cite{hong2020more}&\cite{hong2020deep} &\cite{mohla2020fusatnet}&\cite{jia2023collaborative}&(Raw Data)&{(Augmented Data)} \\
\hline 
OA &0.9474 &0.9841 & 0.9873 & 0.8694 & 0.9770 & {0.9917}&\textcolor{red}0.9949 &\textcolor{blue}{0.9957}\\

AA& 0.9298& 0.9695&0.9763 & 0.8525 &0.9555 &0.9841& \textcolor{red}{0.9915}&\textcolor{blue}{0.9934}\\
kappa &0.9299 &0.9788 & 0.9831 & 0.8255 & 0.9693 &{0.9890}& \textcolor{red}0.9927&\textcolor{blue}{0.9943}\\
\hline
\hline 
\end{tabular}
\label{trento_fusion}
\end{table*}

\subsubsection{\textbf{Results on the MUUFL data set}}
The comparative results from the MUUFL dataset in Table~\ref{mf_fusion} clearly illustrate the effectiveness of our band selection and fusion method. Our method with the raw training data showcases impressive accuracy, while with method with the augmented training data emerges as the leader, outperforming all other models. Specifically, our method trained with the augmented dataset has achieved the highest OA of 0.9957, AA of 0.9934, and Kappa of 0.9943, confirming its superior classification capabilities when utilising a curated selection of 10 bands, as opposed to the full-band approach of the competing models.

\begin{table*}[htp!]
\centering
\caption{Classification performance on the MUUFL data set. All other bands used all bands in data fusion, while our method used only 10 selected bands. The best results are highlighted using \textcolor{blue}{blue color} and the second best results are highlighted using \textcolor{red}{red color.}}
\fontsize{9}{11}\selectfont
\begin{tabular}{|c||c||c||c||c||c||c||c||c|}
\hline 
\hline 
Metrics & CNN-HSI & CoupledCNN& Middle\_Fusion  & EndNet& FusAtNet & CCL & Our Method& Our Method\\
&\cite{yu2017convolutional}&\cite{hang2020classification} &\cite{hong2020more}&\cite{hong2020deep} &\cite{mohla2020fusatnet}&\cite{jia2023collaborative}&(Raw Data)&{(Augmented Data)} \\
\hline 
OA &0.7384&0.7743&0.7806 & 0.7915 & 0.7076& 0.8111& \textcolor{red}{0.9192}&\textcolor{blue}{0.9240}\\
AA & 0.7626&0.7640 &0.7697&0.7852&0.6773& 0.7800 & \textcolor{red}{0.9283}&\textcolor{blue}{0.9306}\\
Kappa &0.6748&0.7133 & 0.7217 & 0.7366 & 0.6377 & 0.7562 & \textcolor{red}{0.8947}&\textcolor{blue}{0.9006}\\
\hline
\hline 
\end{tabular}
\label{mf_fusion}
\end{table*}

All three experiments demonstrate the mechanism of cross-attention in discerning relevant HSI bands that are correlated with LiDAR. This approach shows the potential to enhance remote sensing applications through intelligent band selection guided by LiDAR data, and further benefit for fusion model development.

\section{Conclusions} \label{sec:conclusions}
In this paper, we have introduced a method for hyperspectral band selection guided by the LiDAR data. Our method integrates band selection with the fusion model and has demonstrated superior classification performance compared to full-band fusion models. Specifically, it has shown a notable increase in the overall accuracy, average accuracy, and Kappa coefficients across multiple data sets. This indicates a robust capability to discern and use the most informative spectral bands, which are paired with LiDAR, thus optimising the balance between data volume and classification performance. By selecting bands first and then applying the fusion model, our approach presents a viable solution for real-time processing applications where speed and accuracy are critical. Our future research will investigate the potential of transfer learning to generalise the application of band selection with hyperspectral and LiDAR data fusion. We plan to test the model in a real-time processing environment to evaluate its effectiveness and efficiency in operational scenarios.  




\bibliographystyle{IEEEtran}

\bibliography{IEEEabrv, HSI_Lidar_TGRS.bib}

\begin{IEEEbiography}[{\includegraphics[width=1in,height=4in,clip,keepaspectratio]{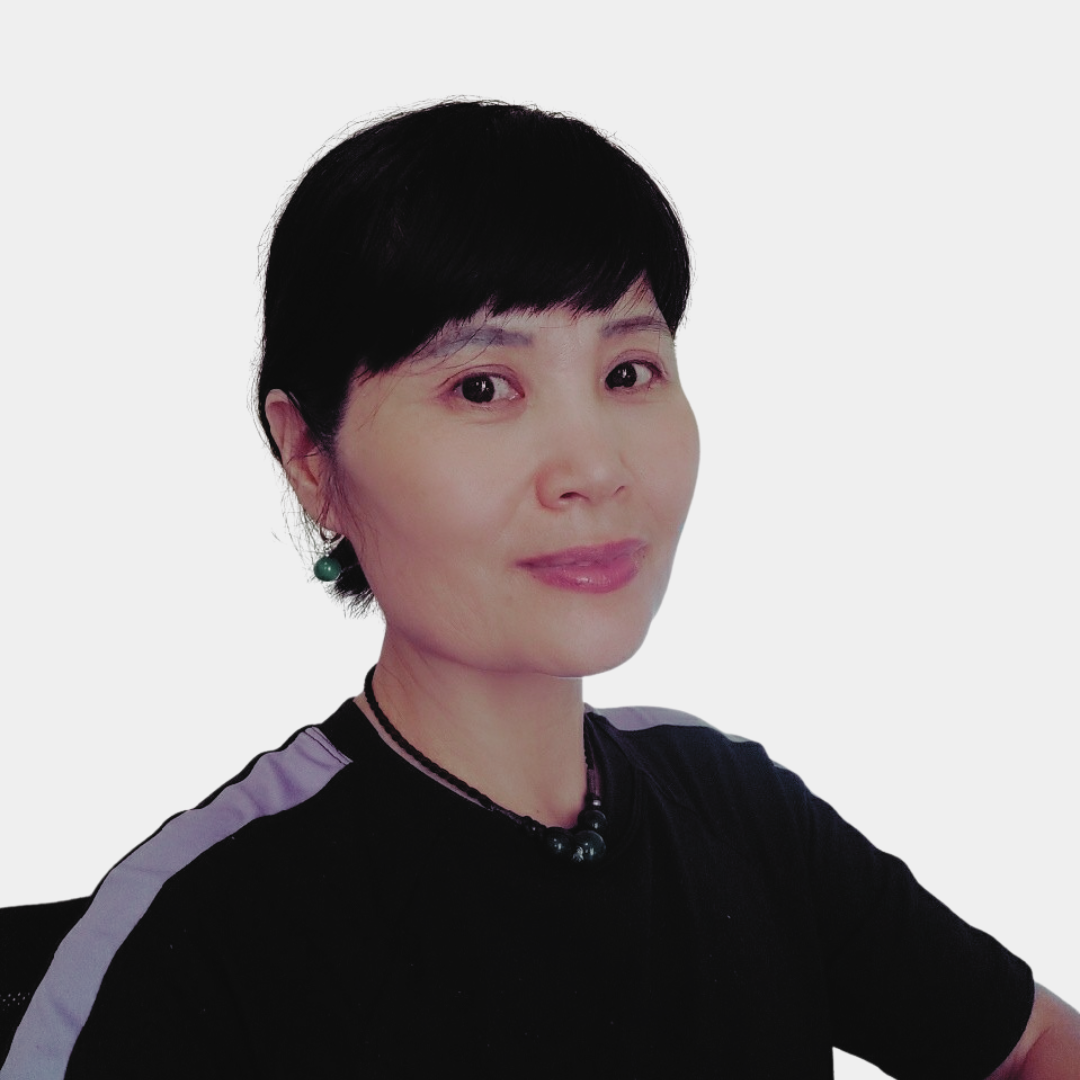}}]{Judy X YANG} (Student Member, IEEE)  worked in manufacturing engineering and supply chain management for more than 15 years as an engineer. Her work history includes in an American manufacturing company PennEngineering in Shanghai as a mechanical project engineer and MSI Electrical Company as an OEM product manager. She has been a member of Engineers Australia since 2015. She completed her Master of Informatics at Central Queensland University, Australia, in 2021.  She is working toward a Ph.D. at Griffith University, Nathan, QLD, Australia. Her research interests include data analysis, Machine Learning, Computer Vision, Hyperspectral image analysis, LiDAR analysis, and multisource remote sensing data fusion.
\end{IEEEbiography}

\begin{IEEEbiography} [{\includegraphics[width=1in,height=4in,clip,keepaspectratio]{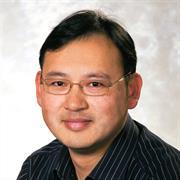}}]{Jun ZHOU} Jun Zhou (Senior Member, IEEE) received a B.S. degree in computer science and a B.E. degree in international business from Nanjing University of Science and Technology, Nanjing, China, in 1996 and 1998, respectively, the M.S. degree in computer science from Concordia University, Montreal, QC, Canada, in 2002, and the Ph.D. degree in computer science from the University of Alberta, Edmonton, AB, Canada, in 2006. He is currently a Professor with the School of Information and Communication Technology, Griffith University, Nathan, QLD, Australia. Previously, he had been a Research Fellow with the Research School of Computer Science, Australian National University, Canberra, ACT, Australia, and a Researcher with the Canberra Research Laboratory, National ICT Australia, Sydney, NSW, Australia. His research interests include pattern recognition, computer vision, and spectral imaging with applications in remote sensing and environmental informatics.\end{IEEEbiography}

\begin{IEEEbiography} [{\includegraphics[width=1in,height=4in,clip,keepaspectratio]{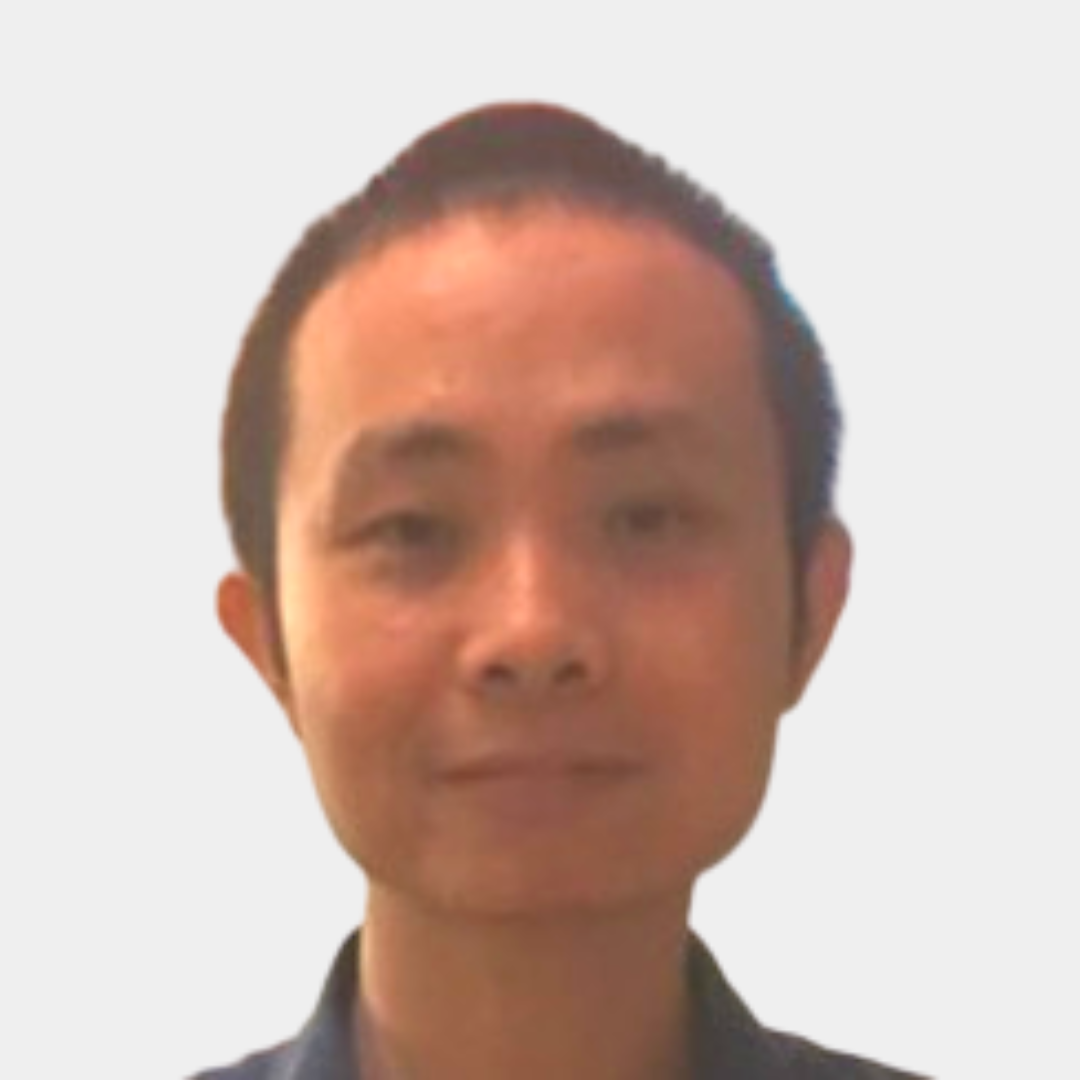}}]{Jing, WANG} Jing Wang received the bachelor’s degree in communication engineering from the Tianjin University of Science and Technology, Tianjin, China, in 2010 and the master’s degree in signal and information processing from the Chinese Academy of Sciences of the University of Beijing, China, in 2013. and in 2020, received the Ph.D. degree from Griffith University, Nathan, QLD, Australia and the Chinese Academy of Sciences University. He is now a scientist at the Agriculture Department. His research interests include pattern recognition and machine learning, computer vision, hyperspectral image analysis, and remote sensing.\end{IEEEbiography}

\begin{IEEEbiography} [{\includegraphics[width=1in,height=4in,clip,keepaspectratio]{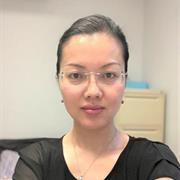}}]{Hui TIAN} Assoc Prof. Hui Tian is Discipline Head of Computer Science in the School of Information and Communication Technology, Griffith, Australia. She received the PhD degree in Computer Science from Japan Advanced Institute of Science and Technology. Her main research interests include Network Routing and Tomography, Privacy-preserving Computing and Knowledge Discovery. She has published one book and more than 80 research papers and led several research projects in different countries. She has actively engaged in professional activities including service as associate editor of SCI-indexed journals and program chair/committee member of international conferences. Assoc Prof. Tian is a senior member of IEEE.
\end{IEEEbiography}

\begin{IEEEbiography} [{\includegraphics[width=1in,height=4in,clip,keepaspectratio]{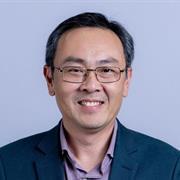}}]{Alan Wee-Chung Liew} PROFESSOR ALAN LIEW is currently the Head of School, the School of Information  Communication Technology,  and the Deputy Director of the Institute for Integrated Intelligent Systems (IIIS), at Griffith University, Australia. Prior to joining Griffith University in 2007, he was an Assistant Professor at the Department of Computer Science and Engineering, Chinese University of Hong Kong. His research interest is in the field of AI, machine learning, medical imaging, computer vision, and bioinformatics. He is leading the medical/health informatics stream in IIIS, and is engaging with the Gold Coast University Hospital and Queensland Health on several collaborative projects. He has also engaged actively in professional activities such as on the organizing committee or technical program committee of many conferences, on editorial boards, as assessor for nationally competitive research grants, and reviewers for many international conferences and journals. He is currently serving as an associate editor of IEEE Trans. Fuzzy Systems, Springer Nature Computer Science, and International Journal of Computational Intelligence Systems, Machine Intelligence Research. He is a senior member of IEEE since 2005.\end{IEEEbiography}

\vfill

\end{document}